\documentclass[superscriptaddress,aps,letterpaper,10pt,twocolumn,floats,showpacs,amsmath,amsfonts,amssymb,prx]{revtex4-2}

\usepackage{graphicx}
\usepackage[colorlinks=true,citecolor=blue]{hyperref}
\usepackage[caption=false]{subfig}
\usepackage{pstricks}
\usepackage{pst-node}
\usepackage{amsmath}
\usepackage{amsbsy}
\usepackage{verbatim}
\usepackage{dsfont}
\begin{document}
		
\title{Quantum and classical contributions to entropy production in fermionic and bosonic Gaussian systems}
	
\author{Krzysztof Ptaszy\'{n}ski}
\email{krzysztof.ptaszynski@uni.lu}
\affiliation{Complex Systems and Statistical Mechanics, Department of Physics and Materials Science, University of Luxembourg, L-1511 Luxembourg, Luxembourg}
\affiliation{Institute of Molecular Physics, Polish Academy of Sciences, Mariana Smoluchowskiego 17, 60-179 Pozna\'{n}, Poland}

\author{Massimiliano Esposito}
\email{massimiliano.esposito@uni.lu}
\affiliation{Complex Systems and Statistical Mechanics, Department of Physics and Materials Science, University of Luxembourg, L-1511 Luxembourg, Luxembourg}
	
\date{\today}
	
\begin{abstract}
As previously demonstrated, the entropy production -- a key quantity characterizing the irreversibility of thermodynamic processes -- is related to generation of correlations between degrees of freedom of the system and its thermal environment. This raises the question of whether such correlations are of a classical or quantum nature, namely, whether they are accessible through local measurements on the correlated degrees of freedom. We address this problem by considering fermionic and bosonic Gaussian systems. We show that for fermions the entropy production is mostly quantum due to the parity superselection rule which restricts the set of physically allowed measurements to projections on the Fock states, thus significantly limiting the amount of classically accessible correlations. In contrast, in bosonic systems a much larger amount of correlations can be accessed through Gaussian measurements. Specifically, while the quantum contribution may be important at low temperatures, in the high temperature limit the entropy production corresponds to purely classical position-momentum correlations. Our results demonstrate an important difference between fermionic and bosonic systems regarding the existence of a quantum-to-classical transition in the microscopic formulation of the entropy production. They also show that entropy production can be mainly caused by quantum correlations even in the weak coupling limit, which admits a description in terms of classical rate equations for state populations, as well as in the low particle density limit, where the transport properties of both bosons and fermions converge to those of classical particles.
\end{abstract}
	
\maketitle
	
\section{Introduction}

Second law of thermodynamics is one of the most fundamental principles of physics that determines the direction of spontaneous thermodynamic processes. Mathematically, this law is formulated by defining a quantity called the entropy production $\sigma$ and postulating that for every thermodynamic process this quantity is nonnegative: $\sigma \geq 0$. Entropy production is thus a key measure of irreversibility of thermodynamic processes; see Refs.~\cite{seifert2012, landi2021} for a review of basic issues related to the description and applications of the entropy production in both classical and quantum setups.
	
Since the time of Boltzmann and his famous controversy with Loschmidt, one of the most basic goals of statistical physics has been to justify the emergence of thermodynamic irreversibility from time-reversal symmetric laws of classical or quantum mechanics. Recent decades have brought about important progress in this area. The explanations for the emergence of irreversibility can be divided into two classes. The first set of arguments rationalizes the relaxation of closed quantum systems to thermal equilibrium by employing concepts such as typicality~\cite{tasaki1998, goldstein2006, popescu2006, baldovin2021, reimann2022}, nonintegrability~\cite{banuls2011} or eigenstate thermalization hypothesis~\cite{deutsch1991, srednicki1994, deutsch2018} (see Refs.~\cite{gogolin2016, polkovnikov2011, eisert2015, alessio2016, mori2018} for reviews). This approach has been supported by experiments~\cite{trotzky2012, clos2016, kaufman2016, neill2016} and may be generalized to demonstrate the emergence of nonequilibrium steady states~\cite{xu2022}. While such explanations appear to be most fundamental, a non-less important insight into the origin of the second law of thermodynamics is provided by the second class of arguments, which deals with open quantum systems attached to a thermal environment (or several thermal reservoirs) to justify their thermalization and the irreversibility of thermodynamic flows. To this class belong arguments based on the semigroup property of Markovian dynamics~\cite{spohn1978} (with recent generalizations to the non-Markovian case~\cite{strasberg2019, rivas2020}), nonequilibrium fluctuation theorems~\cite{andrieux2009, esposito2009, jarzynski2011, campisi2011, campisi2011b}, or resource theory~\cite{horodecki2013, brandao2013, brandao2015}. Finally, our paper focuses on the information-theoretic approach proposed in Ref.~\cite{esposito2010}, which is applicable to a generic open quantum system arbitrarily strongly coupled to the thermal environment (see Ref.~\cite{strasberg2021} for a similar formalism based on the observational entropy and Ref.~\cite{strasberg2021b} for finite-size corrections). Within this formulation, the entropy production is expressed as a sum of two nonnegative information-theoretic quantities: the mutual information quantifying the system-environment correlations and the relative entropy measuring the displacement of the environment from equilibrium. As further shown in our previous work~\cite{ptaszynski2019}, for macroscopic baths composed of many independent degrees of freedom the latter term can be related to the generation of correlations within the environment; thus, the entropy production can be interpreted as a sum of system-environment and intraenvironment correlations.
	
As the entropy production can be related to generation of multipartite correlations within the system-environment ensemble, the natural question appears whether such correlations are of a classical or quantum nature. The related problem of splitting the entropy production into classical and quantum contributions has previously been investigated in the context of a reduced description of system dynamics~\cite{santos2019, camati2019, latune2020, mohammady2020, goes2020}, for closed systems driven by external time-dependent protocols~\cite{francica2019, varizi2020, varizi2021, varizi2022}, as well as for the total system-environment dynamics of open systems in the case of energy-preserving thermal operations~\cite{santos2019}, or in a generic case within the formalism of observational entropy~\cite{strasberg2020}. In all these studies, the quantum contribution to the entropy production has been defined through the change of quantum coherence in the eigenstate basis of the system Hamiltonian (for the reduced dynamics) or the environment Hamiltonian. Here, using the formulation of the entropy production in terms of correlations, we take another path. Following Refs.~\cite{rulli2011, bradshaw2019}, the classical contribution to the entropy production is defined as a maximum amount of correlations accessible through local measurements on the degrees of freedom of the system and the environment. Accordingly, the inaccessible part of correlations is referred to as multipartite (or global) quantum discord. We note that while quantum and classical correlations~\cite{aguilar2022} as well as entanglement~\cite{aguilar2020} between two degrees of freedom of the environment have recently been investigated, our article deals with genuine multipartite correlations. 
	
Specifically, we focus on two types of physical setups: Gaussian fermionic and bosonic systems described by quadratic Hamiltonians. Such systems admit a convenient description since their properties can be characterized by means of two-point correlation matrices instead of much larger-dimensional density matrices. Using analytic arguments and numerical simulations of the system-environment dynamics, we show that for both fermions and bosons the entropy production is mostly related to generation of quantum coherence in the Fock basis (i.e., basis of states with a defined particle number). At first glance, this might suggest that the entropy production is mostly of a quantum nature. However, there is a crucial qualitative difference between fermions and bosons. For the first class of particles, the Fock basis is the only measurement basis allowed due to the parity superselection rule which prohibits the superpositions of states with different particle parity~\cite{wick1952, szalay2021, faba2021}; therefore, correlations in fermionic systems are indeed mostly quantum. On the other hand, for bosons a much larger amount of correlations can be accessed by performing Gaussian heterodyne measurements. In particular, while quantum correlations may be significant for low temperatures, in the high temperature limit the entropy production corresponds to purely classical correlations between positions and momenta of different bosonic modes (e.g., harmonic oscillators). One may observe, therefore, a quantum-to-classical transition in the microscopic formulation of the entropy production, which is absent in the fermionic case. Furthermore, we demonstrate that quantum system-environment and intraenvironment correlations can be present even when the reduced dynamics of the system is apparently classical (i.e. lack of coherence in the system eigenbasis) or in the limit of low particle density where the macroscopic transport properties of both fermions and bosons converge to those of classical particles.
	
The paper is organized as follows. In Sec.~\ref{sec:entrcor} we briefly review the formalism relating the entropy production to correlations and discuss the decomposition of correlations into classical and quantum contributions. In Secs.~\ref{sec:ferm}--\ref{sec:bos} we present the evaluation of quantum and classical contributions to the entropy production in fermionic and bosonic systems, respectively. In Sec.~\ref{sec:lowden} we discuss the microscopic nature of the entropy production for fermions and bosons in the low density limit. Finally, in Sec.~\ref{sec:concl} we present conclusions following from our results. The Appendixes~\ref{sec:hsnorm}--\ref{sec:covev} contain supplementary results, certain derivations, and technical details of calculations.
	
\section{Entropy production as correlation} \label{sec:entrcor}
Let us first present a general formalism relating the entropy production to the generation of correlations. First, up to Eq.~\eqref{denv}, we will review the main results of Refs.~\cite{esposito2010, ptaszynski2019}; next we will discuss the decomposition of correlations into classical and quantum contributions.
	
We consider a generic open quantum system described by the Hamiltonian
\begin{align}
\hat{H}_{SE}=\hat{H}_{S}+\hat{H}_{E}+\hat{H}_{I},
\end{align}
where $\hat{H}_{S}$, $\hat{H}_{E}$, and $\hat{H}_{I}$ are the Hamiltonians of the system, the environment, and the interaction between them, respectively. We will further take the environment to consist of $\mathcal{M}$ baths $\alpha$, each composed of $K_\alpha$ independent degrees of freedom $k$ (later referred to as modes):
\begin{align} \label{hamassum}
\hat{H}_E=\sum_\alpha \sum_{k=1}^{K_\alpha} \hat{H}_{\alpha k}.
\end{align}
Such an assumption is justified, e.g., for baths composed of noninteracting bosons or fermions. As a matter of fact, as shown in condensed matter theory, the effective degrees of freedom of interacting systems can often be represented as noninteracting quasiparticles, such as phonons or magnons. For the sake of simplicity, from hereon we will focus on a single bath case with $K_\alpha=K$, $\hat{H}_{\alpha k}=\hat{H}_k$, etc.; all formulas can be easily generalized to a multiple bath scenario considered in Sec.~\ref{sec:lowden}.
	
The system-environment ensemble is taken to form a closed quantum system undergoing a unitary evolution given by the von Neumann equation 
\begin{align} \label{vnm}
	i \frac{d}{dt} \rho_{SE}=\left[ \hat{H}_{SE},\rho_{SE} \right].
\end{align}
From hereon we take $\hbar=1$. The evolution of the system is assumed to start from the initially uncorrelated state ${\rho_{SE}(0)=\rho_S(0) \otimes \rho_E^\text{eq}}$, where $\rho_S(0)$ is an arbitrary state of the system and $\rho_E^\text{eq}$ is the grand canonical Gibbs state of the environment,
\begin{align} \label{gibbs}
	\rho_E^\text{eq}=\bigotimes_k \rho_k^\text{eq},
\end{align}
where
\begin{align}
	\rho_k^\text{eq}=\frac{e^{-\beta (\hat{H}_k-\mu \hat{N}_k)}}{\text{Tr} \left[ e^{-\beta (\hat{H}_k-\mu \hat{N}_k)} \right]}
\end{align}
is the equilibrium state of the mode $k$; here $\beta=1/(k_B T)$ is the inverse temperature of the environment, $\mu$ is the chemical potential, and $\hat{N}_k$ is the particle number operator acting on the mode $k$.

The entropy production during the time interval $[0,t]$ is defined as
\begin{align}
	\sigma=\Delta S_S - \beta Q,
\end{align}
where $\Delta S_S=S_S(t)-S_S(0)$, $S_S=-\text{Tr} (\rho_S \ln \rho_S)$ is the von Neumann entropy of the system, and $Q$ is the heat extracted from the environment defined as
\begin{align} \label{heatdef}
	Q=-\text{Tr} \left\{ \hat{H}_E \left[\rho_E(t)-\rho_E^\text{eq} \right] \right\}+\mu \text{Tr} \left\{ \hat{N}_E \left[\rho_{E}(t)-\rho_{E}^\text{eq} \right] \right\},
\end{align} 
where $\hat{N}_E=\sum_k \hat{N}_k$ is the particle number operator acting on the environment. Here, the first term corresponds to the energy change of the environment (with a minus sign), while the second one corresponds to the chemical work. By inserting Eqs.~\eqref{hamassum} and~\eqref{gibbs} into the formula above, a heat-related contribution to the entropy production can be expressed as
\begin{align}
	-\beta Q=\sum_k \Delta S_k + \sum_k D(\rho_k||\rho_k^\text{eq}),
\end{align}
where $\Delta S_k=S_k(t)-S_k(0)$ is the change in the von Neumann entropy of the mode $k$ and $D(\rho||\sigma)={\text{Tr} [\rho(\ln \rho-\ln \sigma)]}$ is the relative entropy; from hereon we use a shortened notation $\rho_k(t)=\rho_k$, $S_k(t)=S_k$, etc. As a result, the entropy production can be expressed as the sum of two information-theoretic terms,
\begin{align} \label{entrassum}
	\sigma= I_M+D_\text{env},
\end{align}
where
\begin{align}
	I_M=S_S+\sum_k S_k -S_{SE}
\end{align}
is the multipartite mutual information between the system and the modes of the environment (later referred to as the total correlation), while the quantity 
\begin{align} \label{denv}
	D_\text{env}=\sum_k D(\rho_k||\rho_k^\text{eq})
\end{align}
measures the displacement of the environmental modes from equilibrium; in deriving Eq.~\eqref{entrassum} one uses the assumption of an initially uncorrelated state, which implies $I_M(t=0)=0$. According to information theory, both contributions $I_M$ and $D_\text{env}$ are nonnegative, which guarantees the nonnegativity of the entropy production. For large baths (i.e., in the thermodynamic limit), when the environmental modes are infinitesimally displaced from equilibrium, the contribution $D_\text{env}$ is usually negligible (although may still be important when only certain modes of the environment are resonantly excited~\cite{colla2021}) and the entropy production is dominated by the total correlation $I_M$~\cite{ptaszynski2019}.

This raises the natural question of whether the intermode correlations are of a classical or a quantum nature. In order to answer this question, following Refs.~\cite{rulli2011, bradshaw2019} we will decompose the total correlation as
\begin{align} \label{infdecomp}
	I_M=J_M+\mathcal{D}_M,
\end{align}
where $J_M$ is the classical correlation between the system and the modes of the environment, while $\mathcal{D}_M$ is the multipartite quantum discord. The first quantity is defined as the maximum amount of correlations accessible through local measurements:
\begin{align} \label{classcor}
	J_M=\max_{\{\Pi_S\},\{\Pi_1\},\ldots,\{\Pi_K\}} & \left[ H(\mathcal{A}_S)+\sum_{k} H(\mathcal{A}_k) \right. \\ \nonumber & \left. -H(\mathcal{A}_S \mathcal{A}_1 \ldots \mathcal{A}_K) \right],
\end{align}
where $\{\Pi_i\}$ is a set of measurements acting on the subsystem $i \in \{S,k\}$, $H(\mathcal{A}_i)$ is a Shannon entropy of the measurement outcomes $\mathcal{A}_i$ for a single subsystem $i$, and $H(\mathcal{A}_S \mathcal{A}_1 \ldots \mathcal{A}_K)$ is the Shannon entropy of the measurement outcomes for the total system-environment ensemble. 

We note that while (following Refs.~\cite{rulli2011, bradshaw2019}) the classical correlation is here taken to correspond to local measurements on each mode, a less stringent definition can be provided by allowing nonlocal measurements on sets of several modes. We discuss this in more detail in Sec.~\ref{sec:fermnumsecor}.
	
	\section{Fermionic systems} \label{sec:ferm}
	\subsection{Analytic arguments} \label{sec:ferman}
We will now apply our decomposition of the entropy production to specific physical scenarios, starting from the system of noninteracting fermions. Within the formalism of second quantization, a generic noninteracting fermionic system can be described by a quadratic Hamiltonian~\cite{peschel2003, bravyi2005, kraus2009, hackl2021, surace2022, eisler2015}
	\begin{align} \label{hamfermgen}
		\hat{H}=\sum_{ij} \left(\mathcal{A}_{ij} c_i^\dagger c_j+\mathcal{B}_{ij} c^\dagger_i c^\dagger_j-\mathcal{B}_{ij}^* c_i c_j \right), 
	\end{align}
 where $\mathcal{A}_{ij}=\mathcal{A}_{ji}^*$ and $\mathcal{B}_{ij}=\mathcal{B}_{ji}^*$ are complex numbers, while $c_i^\dagger$ and $c_j$ are fermionic creation and annihilation operators. We will further restrict our discussion to the particle number preserving Hamiltonians with $\mathcal{B}_{ij}=0$.

To make a connection with the thermodynamics formalism presented in Sec.~\ref{sec:entrcor}, we will focus on the case where a single mode $i=0$ belongs to the system and $K$ other modes $i \in \{1,\ldots,K\}$ to the environment (generalization to multimode systems is straightforward). Furthermore, we will take the modes of the environment to be uncoupled ($\mathcal{A}_{ij}=0$ for $i\neq j$ unless $i=0$ or $j=0$); in fact, every quadratic Hamiltonian can be brought to such a form by a unitary transformation. Then, the Hamiltonian~\eqref{hamfermgen} can be rewritten as
	\begin{align} \label{hamnrl}
		\hat{H}_{SE}=\epsilon_0 c^\dagger_0 c_0 +\sum_{i=1}^K \epsilon_{i} c_i^\dagger c_i + \sum_{i=1}^K \left( t_i c^\dagger_0 c_i + \text{h.c.} \right),
	\end{align}
where $\epsilon_i$ are the mode energies and $t_i$ are the tunnel couplings between the system and the modes of the environment. The system and the environment will be further assumed to be initialized in thermal states of their respective Hamiltonians, which belong to the class of fermionic Gaussian states. Then, the evolution generated by an arbitrary (possibly time-dependent) Hamiltonian preserves the Gaussianity of the system-bath state (as well as state of any subsystem). As shown by Peschel~\cite{peschel2003}, the properties of fermionic Gaussian states (for particle number preserving Hamiltonians) can be fully characterized by a correlation matrix $\mathcal{C}$ with matrix elements defined as
	\begin{align}
		\mathcal{C}_{ij}=\langle c_i^\dagger c_j \rangle=\text{Tr} (c^\dagger_i c_j \rho_{SE}).
	\end{align}
    (For a more generic Hamiltonian~\eqref{hamfermgen} with $\mathcal{B}_{ij} \neq 0$ one needs also to consider correlations $\langle c_i c_j \rangle$ and $\langle c_i^\dagger c_j^\dagger \rangle$~\cite{bravyi2005, kraus2009, hackl2021, surace2022, eisler2015}.) Accordingly, the reduced state of any portion of the global system (e.g., system, environment, or a single mode) is described by a corresponding submatrix of the total correlation matrix. In particular, the von Neumann entropy of a system described by the correlation matrix $\mathcal{C}$ (which is required to calculate the total correlation $I_M$) can be calculated as~\cite{sharma2015}
	\begin{align} \label{vnmcorf}
		S=\sum_n \left[ -g_n \ln g_n - (1-g_n) \ln (1-g_n) \right],
	\end{align}
	where $g_n$ are the eigenvalues of the correlation matrix.
	
	We will now argue, based on analytic arguments, that for large environments the total correlation $I_M$ -- and thus the entropy production -- is mostly determined by the quantum discord $\mathcal{D}_M$, while the classical correlation $J_M$ is negligible (this will be confirmed in Sec.~\ref{sec:fermnum} using numerical simulations). To do so, let us express the correlation matrix of the system-environment ensemble as
	\begin{align}
		\mathcal{C} = \mathcal{C}^D+\epsilon \mathcal{E},
	\end{align}
	where $\mathcal{C}^D$ and $\epsilon \mathcal{E}$ are the diagonal and the off-diagonal parts of the correlation matrix, respectively. We now pose the following hypothesis: For large baths the off-diagonal elements $\epsilon \mathcal{E}_{ij}$ are small, such that one can treat $\epsilon$ as a small perturbation parameter. This can be justified as follows: When one simulates the dynamics for different sizes of the environment, as the number of modes $K$ increases, the entropy production at time $t$ converges to some finite value corresponding to the infinite bath limit~\cite{ptaszynski2019}. At the same time, $I_M$ (and thus $\sigma$) is bounded from below by the inequality derived by Bernigau \textit{et al}.~\cite{bernigau2013}:
		\begin{align} \label{boundim}
			\sigma \geq I_M \geq 2\epsilon^2 \text{Tr} (\mathcal{E}^2) = 2\epsilon^2 \sum_{i,j} |\mathcal{E}_{ij}|^2
		\end{align}
		(see Appendix~\ref{sec:hsnorm} for details). Therefore, as the number of elements in the sum above is proportional to $(K+1)K \approx K^2$, while the value of $\sigma$ is fixed, the magnitude of the off-diagonal elements $\epsilon |\mathcal{E}_{ij}|$ is bounded from above by a factor proportional to $1/K$.
	
	In the next step we will demonstrate that, to the lowest order of $\epsilon$, both the total correlation and the classical correlation can be approximated as sums of two-mode total/classical correlations, i.e.,
	\begin{align} \label{infassum}
		I_M &\approx \sum_{i,j>i} I_{ij} \quad \text{for} \quad \epsilon \rightarrow 0, \\ \label{clinfassum}
		J_M &\approx \sum_{i,j>i} J_{ij} \quad \text{for} \quad \epsilon \rightarrow 0,
	\end{align}
	where the lowest-order contribution to the total and the classical two-mode correlation is of the order $\mathcal{O}(\epsilon^2)$ and $\mathcal{O}(\epsilon^4)$, respectively:
	\begin{align} \label{pairtot}
		I_{ij} &= \mathcal{O}(\epsilon^2), \\ \label{paircl}
		J_{ij} & = \mathcal{O}(\epsilon^4).
	\end{align}
	This implies that for small $\epsilon$ the classical correlation $J_M$ is negligible compared to the total correlation $I_M$. More precisely, since $\epsilon$ is bounded by a factor proportional to $1/K$, it can be expected that the ratio $J_M/I_M$ decays as $1/K^2$.
	
	To show that, let us first consider the total correlation $I_M$. Using second-order nondegenerate perturbation theory, the eigenvalues of the correlation matrix can be calculated as
	\begin{align}
		g_i=\mathcal{C}_{ii}+ \epsilon^2 \sum_{i \neq j} \frac{|\mathcal{E}_{ij}|^2}{\mathcal{C}_{ii}-\mathcal{C}_{jj}}+\mathcal{O}(\epsilon^3).
	\end{align}
	Using Eq.~\eqref{vnmcorf}, for small $\epsilon$ the total correlation reads then
	\begin{align} \label{pertitot}
		I_M \approx \sum_{i,j \neq i} \frac{\epsilon^2}{2} \frac{|\mathcal{E}_{ij}|^2}{\mathcal{C}_{ii}-\mathcal{C}_{jj}} \ln \frac{\mathcal{C}_{ii}(1-\mathcal{C}_{jj})}{\mathcal{C}_{jj}(1-\mathcal{C}_{ii})}.
	\end{align} 
	Quite obviously, the contribution linear in $\epsilon$ has to disappear since the mutual information is nonnegative, and thus the lowest-order contribution is quadratic in $\epsilon$. A single element of the sum can easily be identified as the lowest-order contribution to the quantum mutual information between two modes $I_{ij}$ (one simply applies the formula above to a two-mode system), which leads to Eqs.~\eqref{infassum} and~\eqref{pairtot}. As demonstrated numerically in Sec.~\ref{sec:fermnumcor}, formula~\eqref{pertitot} is actually not exact even in the large bath limit; this is related to the presence of redundant correlations, due to which $I_M$ is not exactly equal to $\sum_{i,j>i} I_{ij}$. Nevertheless, it still very well describes the magnitude of $I_M$, which supports our general argument.
	
	Let us now turn our attention to the classical correlation $J_M$. In general, its calculation requires optimization over all possible sets of measurement operators. For fermionic systems, however, the problem is simplified by the parity superselection rule which forbids the existence of quantum superpositions of states with even and odd numbers of particles~\cite{wick1952, szalay2021}; correspondingly, the projective measurements on such superpositions are not allowed. In particular, the only possible projective measurement operators acting on the $i$th fermionic mode are projections on the empty and the occupied state~\cite{faba2021}:
	\begin{align}
		\Pi_i^0 &=c_i c_i^\dagger, \\
		\Pi_i^1 &=c_i^\dagger c_i.
	\end{align}
	As a result, the classical correlation $J_M$ is equal to the mutual information in the Fock basis (occupation basis) $J_M^F$:
	\begin{align} \label{entrfock}
		J_M=J_M^F =\sum_{\mathbf{n}} p(\mathbf{n}) \ln \frac{p(\mathbf{n})}{\pi(\mathbf{n})},
	\end{align}
	where $\mathbf{n}=(n_0,\ldots,n_K)$ is the vector of the mode occupancies, $\pi(\mathbf{n})=\prod_i p(n_i)$, and $\sum_\mathbf{n} = \sum_{n_0} \ldots \sum_{n_K}$.
	
	The probability distribution $p(\mathbf{n})$ is fully determined by its transform, the moment generating function
	\begin{align}
		M(\pmb{\lambda}) =\sum_{\mathbf{n}} p(\mathbf{n}) e^{\pmb{\lambda} \cdot \mathbf{n}}=\sum_{\mathbf{k}} \left \langle \mathbf{n}^\mathbf{k} \right \rangle \frac{\mathbf{l}^\mathbf{k}}{\mathbf{k}!},
	\end{align}
	where $\mathbf{\lambda}=(\lambda_0,\ldots,\lambda_K)$ is the vector of the counting fields, $\sum_{\mathbf{k}}=\sum_{k_0=0}^\infty \ldots \sum_{k_K=0}^\infty$, $\mathbf{n}^\mathbf{k}=\prod_{i=0}^K n_i^{k_i}$ and $\mathbf{k}!=\prod_{i=0}^K k_i!$. Specifically, the state probabilities can be calculated as
	\begin{align} \label{probfrommom}
		p(\mathbf{n})=\frac{1}{2\pi} \int_0^{2\pi} d\lambda_0 \ldots \int_0^{2\pi} d\lambda_K e^{-i\pmb{\lambda} \cdot \mathbf{n}} M(i\pmb{\lambda}).
	\end{align}
	
The statistical moments of the particle number operator can be expressed as
\begin{align}
\langle \mathbf{n}^\mathbf{k} \rangle = \left \langle \prod_{i=0}^K (c^\dagger_i c_i)^{k_i} \right \rangle.
\end{align}
They can be calculated using Wick's theorem, according to which for Gaussian states all higher-order correlations are sums of products of two-point correlations $\langle c_i^\dagger c_j \rangle$. It can be expressed as~\cite{ropke2013}
	\begin{align} \label{wick}
		\left \langle f_1 \ldots f_N \right \rangle =\sum_P \sigma(P) \langle f_1 f_2 \rangle \ldots \langle f_{N-1} f_N \rangle,
	\end{align}
	where $f_i$ can be either creation or annihilation operators, $N$ is an even number, $\sum_{P}$ is the sum over all distinct permutations of indices, and $\sigma(P)$ is the permutation sign (+/- for the even/odd number of index swaps). As one can easily note, every product of two-point correlations contains an even number of the off-diagonal elements of the correlation matrix. As a result, the moments $\langle \mathbf{n}^\mathbf{k} \rangle$ -- and therefore the probabilities $p(\mathbf{n})$ -- depend only on even powers of $\epsilon$. In particular, the terms of the order $\epsilon^2$ correspond to a single perturbation of indices, and therefore to two-mode correlations:
	\begin{align}
		\langle \mathbf{n}^\mathbf{k} \rangle = \left( \prod_{i=0}^K \langle n_i^{k_i} \rangle \right) \left[1 + \sum_{i,j>i}\epsilon^2 g(k_i,k_j)+\mathcal{O} (\epsilon^4) \right],
	\end{align}
	where
	\begin{align}
		\epsilon^2 g(k_i,k_j) = \frac{\left \langle n_i^{k_i} n_j^{k_j} \right \rangle}{\langle n_i^{k_i} \rangle \langle n_j^{k_j} \rangle}-1.
	\end{align}
	Consequently, using Eq.~\eqref{probfrommom}, to the lowest order of $\epsilon$ many body probabilities $p(\mathbf{n})$ describe only two-mode correlations:
	\begin{align} \label{probexp}
		p(\mathbf{n})=\pi(\mathbf{n}) \left[1-\sum_{i,j>i} \epsilon^2 \Delta(n_i,n_j) +\epsilon^4 \theta(\mathbf{n})+\mathcal{O}(\epsilon^6)\right],
	\end{align}
	where
	\begin{align} \label{defdelta}
		\epsilon^2 \Delta(n_i,n_j)=\frac{p(n_i,n_j)}{p(n_i)p(n_j)}-1,
	\end{align}
	and $\theta(\mathbf{n})$ is the contribution of the order $\mathcal{O}(\epsilon^4)$.
	
	Inserting Eq.~\eqref{probexp} into Eq.~\eqref{entrfock} and expanding into Taylor series one gets
	\begin{align} \label{entrfockexp}
		&J^F_M= \epsilon^2 \sum_{\mathbf{n}} \pi(\mathbf{n}) \sum_{i,j>1} \Delta (n_i,n_j)+\epsilon^4 \sum_{\mathbf{n}} \pi(\mathbf{n}) \theta(\mathbf{n}) \\ \nonumber
		& +\frac{\epsilon^4}{2} \sum_{\mathbf{n}} \pi(\mathbf{n}) \sum_{i,j>i} \Delta (n_i,n_j) \sum_{k_,l>k} \Delta (n_k,n_l)+\mathcal{O}(\epsilon^6).
	\end{align}
	We now use the fact that the introduction of correlations does not affect the single-mode probabilities $p(n_i)$,
	\begin{align} \label{probcons}
		\sum_j p(n_j) \Delta (n_i,n_j)=0,	\end{align}
which can be derived using Eq.~\eqref{defdelta} and the relations $\sum_j p(n_i,n_j)=p(n_i)$ and $\sum_j p(n_j)=1$. We further use the total probability conservation: the sum of all probabilities $\sum_{\mathbf{n}} p(\mathbf{n})=\sum_{\mathbf{n}} \pi(\mathbf{n})=1$ is not affected by introduction of correlations, and thus summing Eq.~\eqref{probexp} over all $\mathbf{n}$ one gets
 \begin{align} \label{probcons2}		\sum_\mathbf{n} \pi(\mathbf{n}) \theta (\mathbf{n})=0.
	\end{align}
	As a result, the first two terms on the right-hand side of Eq.~\eqref{entrfockexp} disappear, whereas the third reduces to
	\begin{align}
		&\sum_{\mathbf{n}} \pi(\mathbf{n}) \sum_{i,j>i} \Delta (n_i,n_j) \sum_{k_,l>k} \Delta (n_k,n_l)= \\ \nonumber
		&\sum_{n_1} p(n_1) \ldots \sum_{n_L} p(n_L) \left[ \sum_{i,j>1} \Delta (n_i,n_j) \sum_{k_,l>k} \Delta (n_k,n_l) \right]= \\ \nonumber
		&\sum_{i,j>1} p(n_i) p(n_j) \Delta^2 (n_i,n_j),
	\end{align}
	where Eq.~\eqref{probcons} and $\sum_i p(n_i)=1$ were used. Finally, for small $\epsilon$ we find
		\begin{align} \label{entrfockexp2}
			J^F_M \approx \frac{\epsilon^4}{2} \sum_{i,j>1} p(n_i) p(n_j) \Delta^2 (n_i,n_j).
		\end{align}
		Again, a single element of the sum can be identified as a classical correlation between two modes $J^F_{ij}$, leading to Eqs.~\eqref{clinfassum} and~\eqref{paircl}. As in the case of $I_M$, we do not expect Eq.~\eqref{entrfockexp2} to be exact even in the large bath limit. This is because the terms of the order $\epsilon^4$ and higher in Wick's theorem~\eqref{wick} may actually still contribute to many-body probabilities $p(\mathbf{n})$ due to their large number. Nevertheless, as shown later in Sec.~\ref{sec:fermnumden}, the equation correctly predicts the order of magnitude of $J^F_M$.
	
	\subsection{Two-mode correlations} \label{sec:ferm2mode}
	\begin{figure}
		\centering
		\includegraphics[width=0.9\linewidth]{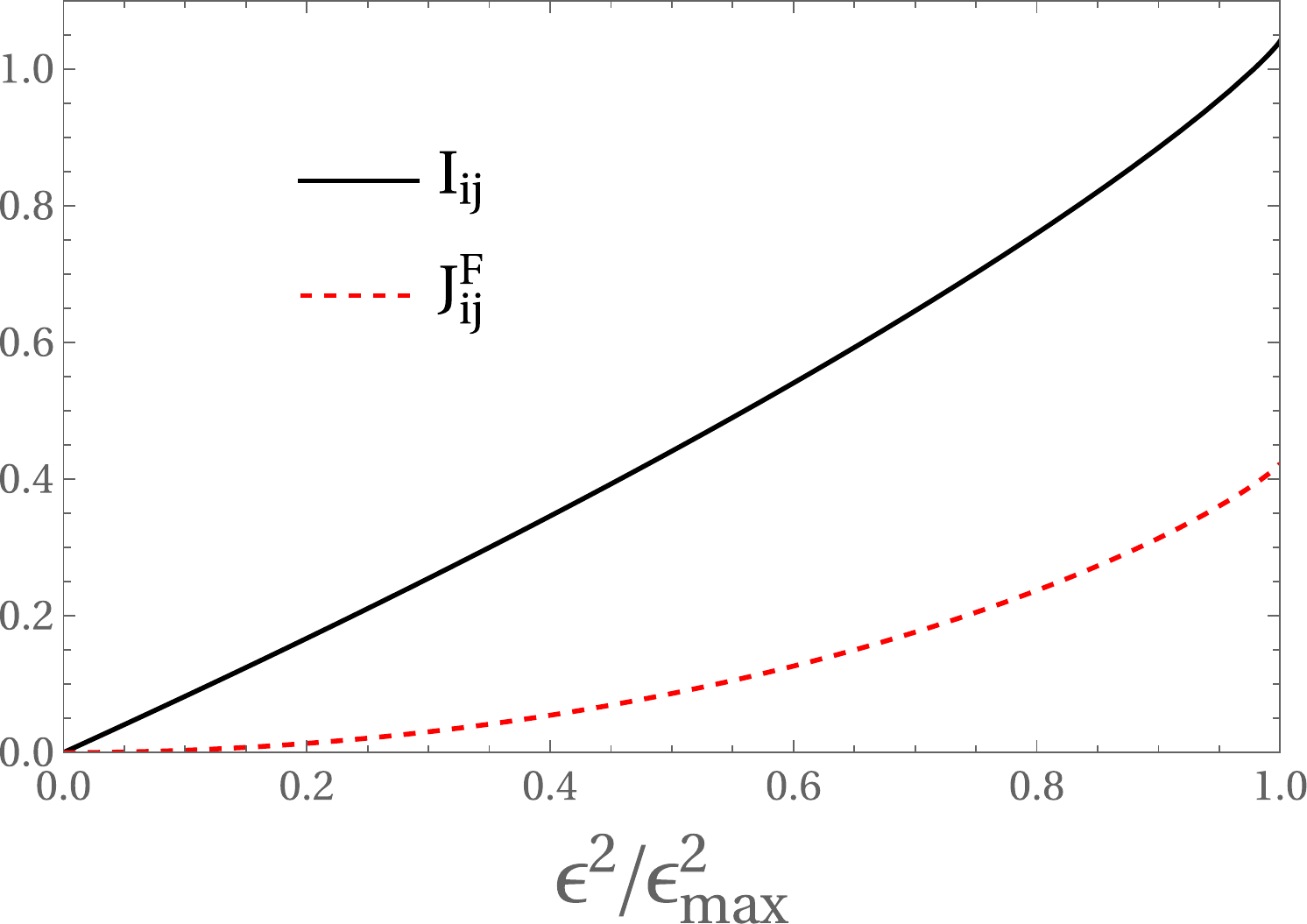}
		\caption{The total mutual information between two fermionic modes $I_{ij}$ and its classical part $J_{ij}^F$ as a function of $\epsilon^2$ for $\langle n_i \rangle=0.5$, $\langle n_j \rangle=0.4$, and $\epsilon_\text{max} = \sqrt{\langle n_i \rangle \langle n_j \rangle}$.}
		\label{fig:ferm2s}
	\end{figure}
	To illustrate the analytic results presented above, let us analyze the total and classical correlations for a pair of modes $i$ and $j$ described by the correlation matrix
	\begin{align} \label{cormat2modferm}
		\mathcal{C}= \begin{pmatrix} \langle n_i \rangle & \epsilon \\ \epsilon & \langle n_j \rangle \end{pmatrix},
	\end{align}
	where $\langle n_i \rangle$ is the average occupancy of the mode $i$ and the correlation term $\epsilon$ is bounded as $\epsilon \leq \epsilon_\text{max} = {\text{min} [\sqrt{\langle n_i \rangle \langle n_j \rangle},\sqrt{(1-\langle n_i \rangle)(1- \langle n_j \rangle)}]}$. Using Wick's theorem~\eqref{wick} the state probabilities $p(n_i,n_j)$ can be calculated as $p(0,0)={(1-\langle n_i \rangle)(1-\langle n_j \rangle)}-\epsilon^2$, $p(0,1)={(1-\langle n_i \rangle) \langle n_j \rangle}+\epsilon^2$, $p(1,0)={\langle n_i \rangle (1-\langle n_j \rangle)}+\epsilon^2$, and $p(1,1)={\langle n_i \rangle \langle n_j \rangle}-\epsilon^2$. The total correlation between the modes and its classical part read as
	\begin{align}
		I_{ij} &=\frac{\epsilon^2}{\langle n_i \rangle- \langle n_j \rangle } \ln \frac{\langle n_i \rangle (1-\langle n_j \rangle)}{\langle n_j \rangle (1-\langle n_i \rangle)}+\mathcal{O}(\epsilon^4) \\
		J_{ij}^F &=\frac{\epsilon^4}{2 \langle n_i \rangle \langle n_j \rangle (1-\langle n_i \rangle)(1-\langle n_j \rangle)} +\mathcal{O}(\epsilon^6).
	\end{align}
	Their magnitude is compared in Fig.~\ref{fig:ferm2s}. As one can observe, for small $\epsilon$ the classical correlation is indeed negligible compared to the total correlation.
	
	\subsection{Numerical results} \label{sec:fermnum}
	\subsubsection{Full density matrix evolution} \label{sec:fermnumden}
	
	\begin{figure}
		\centering
		\includegraphics[width=0.9\linewidth]{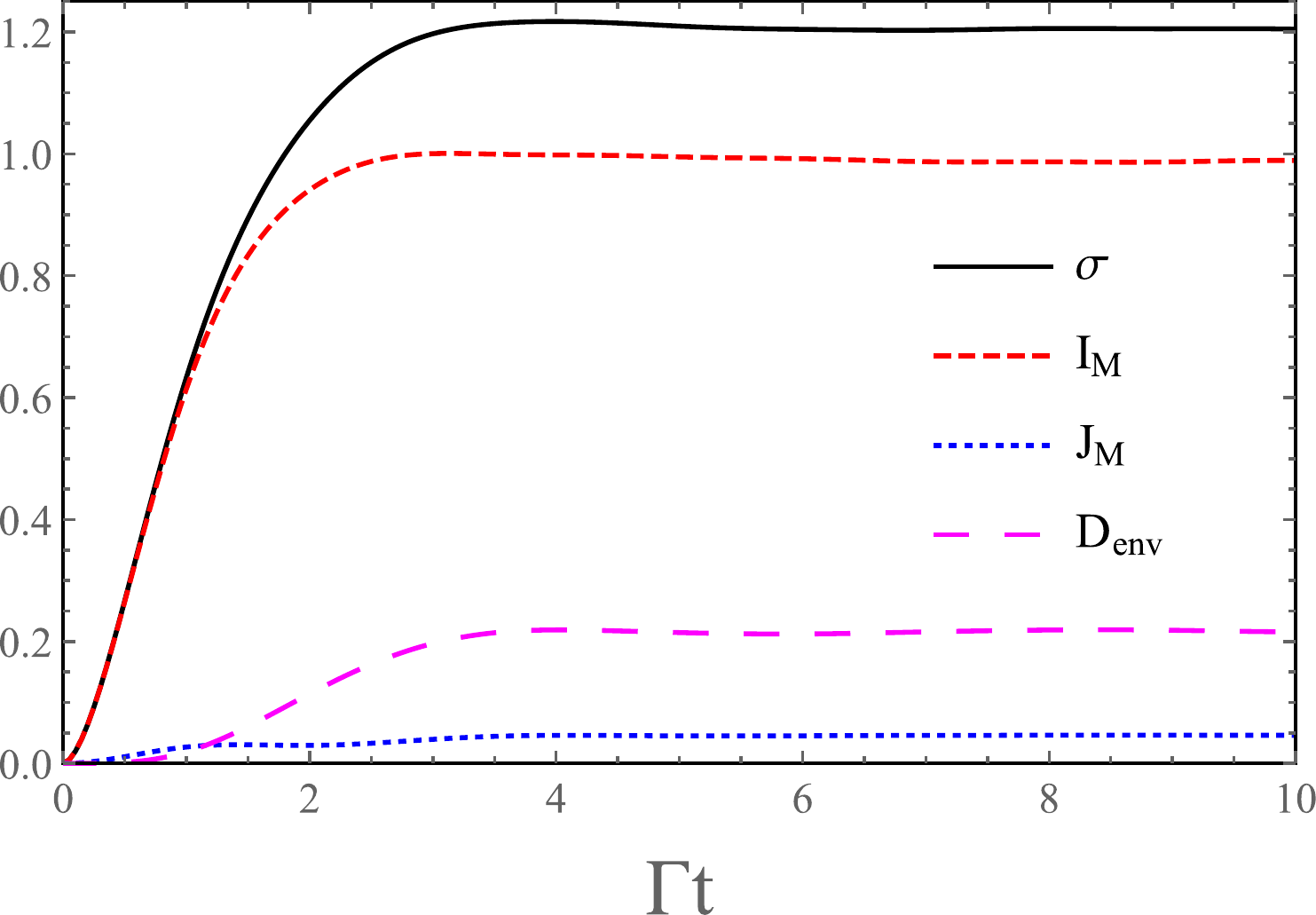}
		\caption{Entropy production and its constituents as a function of time for the initially empty system ($\langle \hat{N}_0(0) \rangle=0$), $\epsilon_0=-0.5 k_B T$, $\Gamma=k_B T$, $\mu=0$, $W=3 k_B T$ and $K=8$.}
		\label{fig:entrprod-ferm}
	\end{figure}
	To further demonstrate the validity of our theoretical reasoning, we perform numerical simulations of the system-environment dynamics for the setup described by the Hamiltonian~\eqref{hamnrl}. The energy levels of the environment are taken to be uniformly distributed over the interval $[-W/2,W/2]$, while the tunnel couplings $t_i$ -- to be equal and parameterized as $t_i=\sqrt{\Gamma W/[2\pi(K-1)]}$, where $\Gamma$ is the coupling strength. In the first step we analyze a global unitary evolution of the system-environment density matrix given by the von Neumann equation~\eqref{vnm}. This approach enables one to calculate the exact value of the classical correlation $J_M$, but is feasible only for baths consisting of just a few modes; however, even for such small reservoirs some relevant results can be obtained. As demonstrated in Fig.~\ref{fig:entrprod-ferm}, already for environments consisting of just 8 modes one can observe a good thermalization behavior. In full agreement with our theoretical predictions, the classical correlation $J_M$ is very small compared to the total correlation $I_M$; therefore, the entropy production is dominated by quantum correlations. At the same time, the contribution $D_\text{env}$, related to the displacement of modes from equilibrium, is here non-negligible; however, as shown later, it can be neglected for larger baths consisting of hundreds of modes.
	
	\begin{figure}
		\centering
		\includegraphics[width=0.9\linewidth]{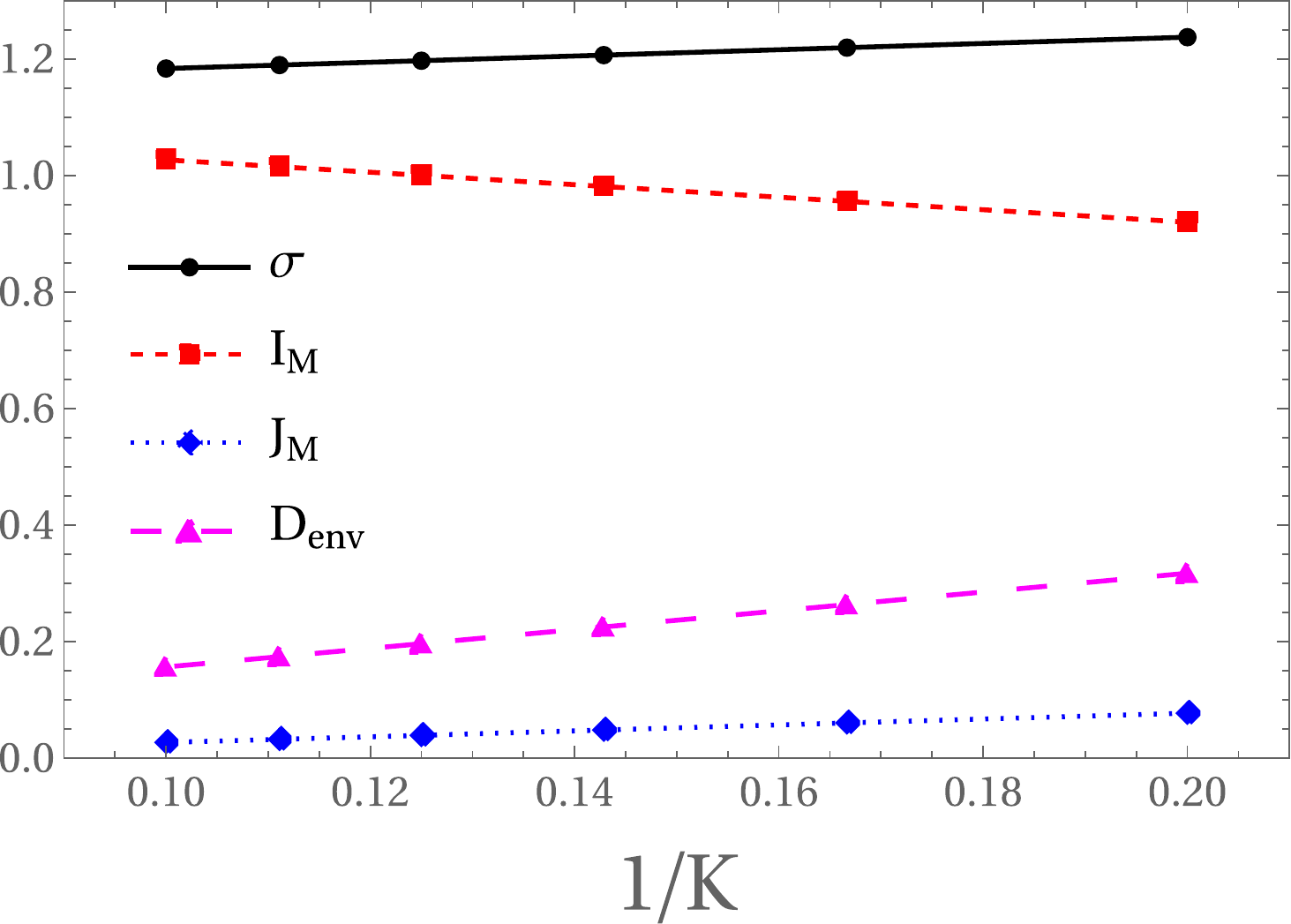}
		\caption{Scaling of the entropy production and its constituents as a function of the inverse of the number of environmental modes $K$ for a fixed time $\Gamma t=3$ and other parameters as in Fig.~\ref{fig:entrprod-ferm}. Lines shown for eye guidance.}
		\label{fig:entrprod-ferm-scaling}
	\end{figure}
	
	A further insight can be provided by analyzing the scaling of the constituents of the entropy production as a function of the size of the environment. As shown in Fig.~\ref{fig:entrprod-ferm-scaling}, both the classical correlation $J_M$ and the displacement term $D_\text{env}$ decrease with bath size, while the total correlation $I_M$ approaches the entropy production $\sigma$. This confirms that for large environments the entropy production is dominated by quantum correlations. More specifically, $D_\text{env}$ scales approximately as $1/K$ while $J_M$ decreases faster than $1/K$ but slower than $1/K^2$, as predicted by the theory presented in Sec.~\ref{sec:ferman}. This discrepancy can be explained by the finite size effects: First, for small baths the approximation $K(K+1) \approx K^2$ for the number of off-diagonal elements of the correlation matrix is not perfectly valid. Second, this approximation is further invalidated by the fact that not all elements $\mathcal{E}_{ij}$ contribute to $I_M$ and $J_M$ with a similar magnitude, but rather the correlations between modes with energy close to $\epsilon_0$ (i.e., resonant with the energy level of the system) are dominant.
	
	\begin{figure}
		\centering
		\includegraphics[width=0.9\linewidth]{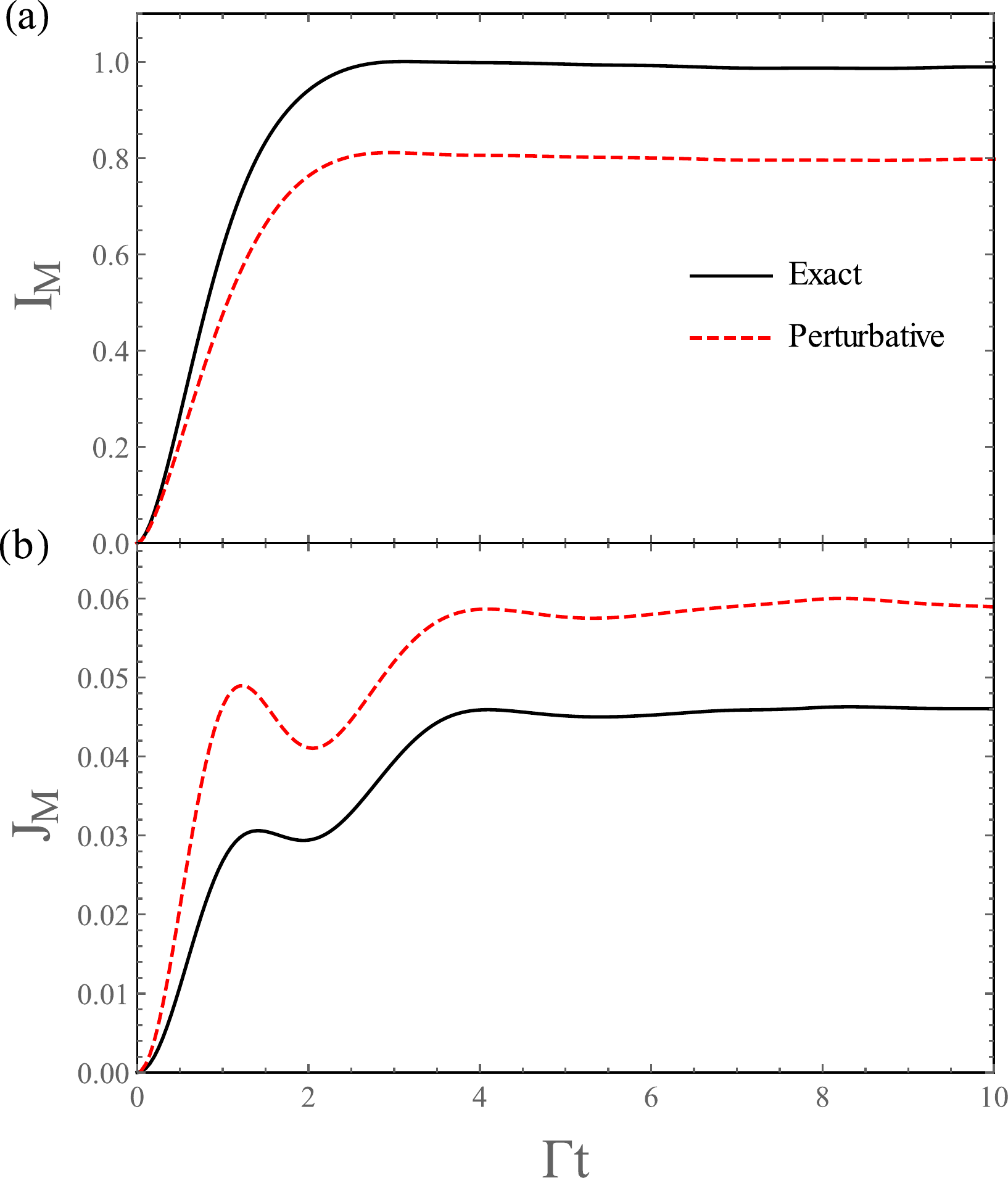}
		\caption{Exact values of the total (a) and the classical (b) correlation (black solid lines) compared with results obtained using perturbative formulas \eqref{pertitot} and \eqref{entrfockexp2} (red dashed lines). Parameters as in Fig.~\ref{fig:entrprod-ferm}.}
		\label{fig:fermcomp}
	\end{figure}
	
	Finally, we check how well the perturbative formulas \eqref{pertitot} and \eqref{entrfockexp2} describe the exact values of $I_M$ and $J_M$. As shown in Fig.~\ref{fig:fermcomp}, even for the small bath analyzed (composed of only 8 modes), they provide results correct to the order of magnitude (though, as later shown, they are not exact even for large baths); this further confirms the validity of analytic arguments presented in Sec.~\ref{sec:ferman}.
	
	\subsubsection{Correlation matrix approach} \label{sec:fermnumcor}
	Due to noninteracting nature of the system, much larger baths (consisting of hundreds of modes) can be analyzed considering the evolution of the correlation matrix $\mathcal{C}$ instead of the density matrix $\rho_{SE}$. The dynamics of the correlation matrix follows the equation~\cite{eisler2012}
	\begin{align}
		\mathcal{C}(t)=e^{i\mathcal{H}t} \mathcal{C}(0) e^{-i\mathcal{H}t},
	\end{align} 
	where $\mathcal{H}$ is the single-particle Hamiltonian defined as
	\begin{align} \label{fermhamsp}
		\begin{cases}
			\mathcal{H}_{ii}= \epsilon_i & \text{for} \quad i=0,\ldots,K, \\
			\mathcal{H}_{0i}=\mathcal{H}_{i0}=t_i & \text{for} \quad i=1,\ldots,K, \\
			\mathcal{H}_{ij}=0 &  \text{otherwise},
		\end{cases}
	\end{align}
	and the initial correlation matrix reads $\mathcal{C}(0)={\text{diag}[0,f(\epsilon_1),\ldots,f(\epsilon_K)]}$, where ${f(\epsilon)=1/\{1+\exp[\beta(\epsilon-\mu)]\}}$ is the Fermi distribution. The particle number and the energy of the bath, which appear in the definition of heat [Eq.~\eqref{heatdef}], can be calculated as
	\begin{align}
		\langle \hat{N}_E \rangle &= \sum_{i=1}^K \mathcal{C}_{ii}, \\
		\langle \hat{H}_E \rangle &= \sum_{i=1}^K \epsilon_i \mathcal{C}_{ii},
	\end{align}
while information-theoretic quantities can be obtained using Eq.~\eqref{vnmcorf}.

\begin{figure}
	\centering
	\includegraphics[width=0.9\linewidth]{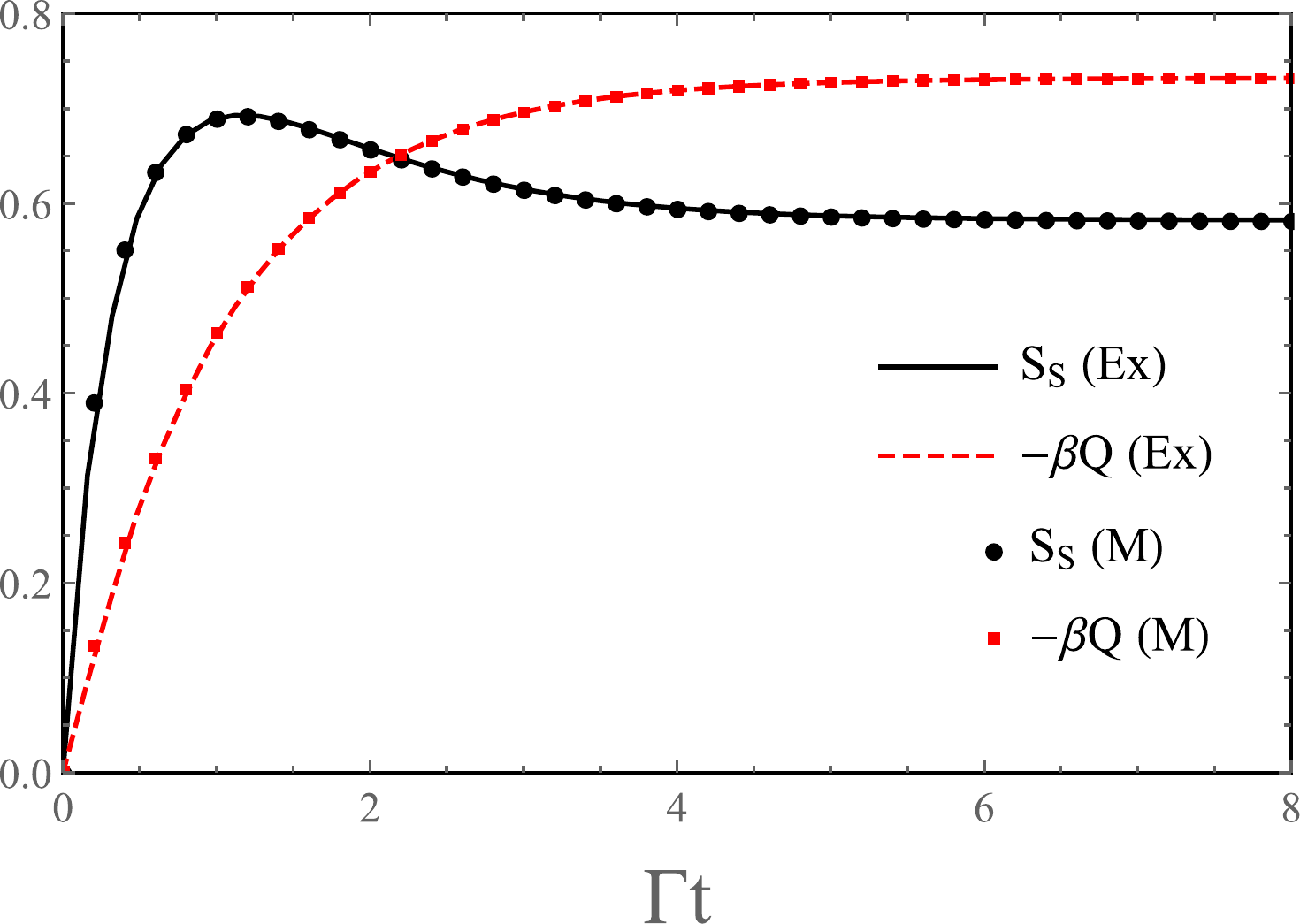}
	\caption{Comparison of evolution of the von Neumann entropy of the system $S_S$ and heat-related contribution to entropy production $-\beta Q$ for the exact (Ex) and Markovian (M) dynamics. Results obtained for the initially empty system ($\langle \hat{N}_0(0) \rangle=0$), $\Gamma=0.02 k_B T$, $\epsilon_0=0$, $\mu=k_B T$, $W=k_B T$ and $K=400$.}
	\label{fig:cormatdyn}
\end{figure}

We will now demonstrate that the quantum contribution to the entropy production dominates even in the regime when the reduced dynamics of the system can be well described by means of a Markovian master equation for states populations. To do so, we will first show that for a weak system-bath coupling $\Gamma \ll k_B T$ the correlation matrix approach gives results that coincide with those provided by the master equation~\cite{fichetti1998}
\begin{align}
	\frac{d}{dt} \langle \hat{N}_0 \rangle = \Gamma \left[f(\epsilon_0)- \langle \hat{N}_0 \rangle \right].
\end{align}
Within this formulation, the heat can be calculated as $Q= {\langle \Delta \hat{N}_0 \rangle (\mu-\epsilon_0)}$, where $\langle \Delta \hat{N}_0 \rangle={\langle \hat{N}_0 (t)\rangle}-{\langle \hat{N}_0(0) \rangle}$. The results for small $\Gamma=0.02 k_B T$ are presented in Fig.~\ref{fig:cormatdyn}; as one can observe, the correlation matrix and the master equation approaches are indeed in almost perfect agreement.

Next, let us analyze the information-theoretic constituents of the entropy production. Within the correlation matrix approach a direct calculation of the classical correlation $J_M$ is not feasible, since this still requires many-body probabilities; though in principle they can be calculated using Wick's theorem~\eqref{wick}, this is computationally difficult. However, one can provide an upper bound. It can be obtained by expressing $J_M$ through the chain rule for mutual information~\cite{rulli2011, bradshaw2019}
\begin{align} \label{chain}
	J_M \leq \sum_{k=0}^{K-1} J_{0 \ldots k,k+1},
\end{align}
where $J_{0 \ldots k,k+1}$ is the classical correlation between the system consisting of modes $0 \ldots k$ and the mode $k+1$ (equality holds if $J_M$ and $J_{0 \ldots k,k+1}$ are optimized in the same measurement basis). The latter quantity obeys the Holevo bound~\cite{holevo1973}
\begin{align} \label{holevo}
	&	J_{0 \ldots k,k+1} \leq \text{max} J_{0 \ldots k,k+1} \\ \nonumber
	&= S_{0 \ldots k}-(1-\langle \hat{N}_{k+1} \rangle) S(\rho^0_{0 \ldots k})-\langle \hat{N}_{k+1} \rangle S(\rho^1_{0 \ldots k}),
\end{align}
where $S(\rho^{0/1}_{0 \ldots k})$ is the von Neumann entropy of the conditional state $\rho^{0/1}_{0 \ldots k}$ given the empty/occupied state of the mode $k+1$; it can be calculated using the fact that conditional states are Gaussian states~\cite{bravyi2005} with correlation matrix elements
\begin{align} \label{cormatcond0}
\mathcal{C}_{ij}^0 &=\mathcal{C}_{ij}+(1- \langle \hat{N}_{k+1} \rangle)^{-1} \mathcal{C}_{i,k+1} \mathcal{C}_{k+1,j}, \\ \label{cormatcond1}
\mathcal{C}_{ij}^1 &= \mathcal{C}_{ij}-\langle \hat{N}_{k+1} \rangle^{-1} \mathcal{C}_{i,k+1} \mathcal{C}_{k+1,j},
\end{align}
which can be derived using Wick's theorem~\eqref{wick} (see Appendix~\ref{sec:cormatcond} for details). Thus, finally,
\begin{align} \label{chainrule}
	J_M \leq \text{max} J_M = \sum_{k=0}^{K-1} \text{max} J_{0 \ldots k,k+1}.
\end{align}

\begin{figure}
	\centering
	\includegraphics[width=0.9\linewidth]{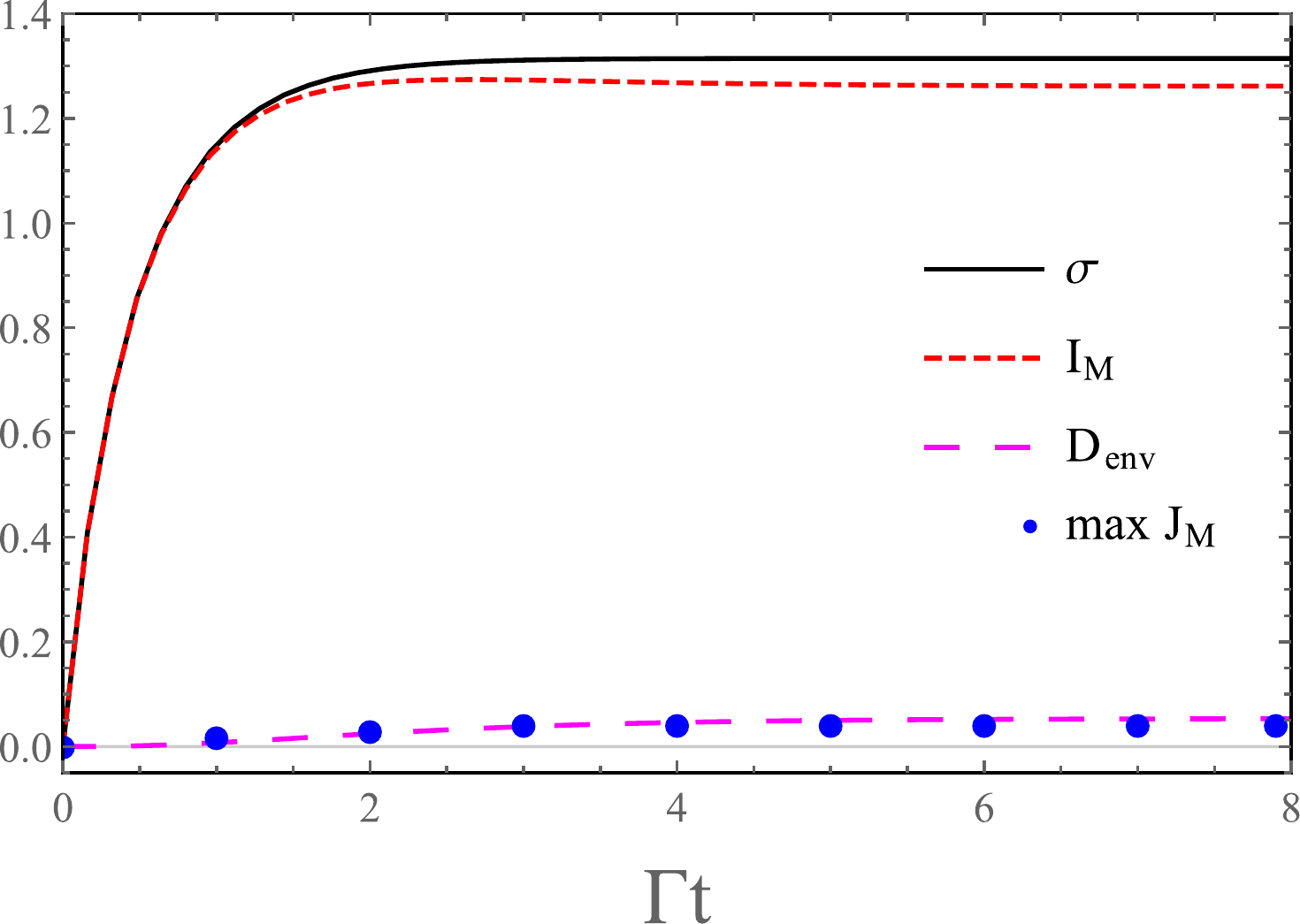}
	\caption{Entropy production and its constituents as a function of time for parameters as in Fig.~\ref{fig:cormatdyn}.}
	\label{fig:cormatentr}
\end{figure}
The results are presented in Fig.~\ref{fig:cormatentr}. They show that the entropy production is clearly dominated by quantum correlations (as classical contributions $J_M$ and $D_\text{env}$ are negligible), even though the reduced dynamics and thermodynamics of the system are well described by a classical Markovian rate equation for the state populations. This resembles a similar previous observation for the pure dephasing case: The character of the reduced dynamics of the system is not directly related to the quantum or classical character of the system-environment and intraenvironment correlations, since by performing the partial trace one loses track of details of the microscopic dynamics~\cite{pernice2012, smirne2021}. Furthermore, this clearly implies that classical and quantum microscopic correlations are fundamentally distinct from classical and quantum contributions to entropy production defined in Refs.~\cite{santos2019, camati2019, latune2020, mohammady2020} for the reduced dynamics. Indeed, in these works quantum contribution to entropy production was related to the dynamics of off-diagonal elements in the system density matrix (in the eigenbasis of the system Hamiltonian). Thus, it vanishes for systems described by classical rate equations, in contrast to microscopic quantum correlations.

\begin{figure}
	\centering
	\includegraphics[width=0.9\linewidth]{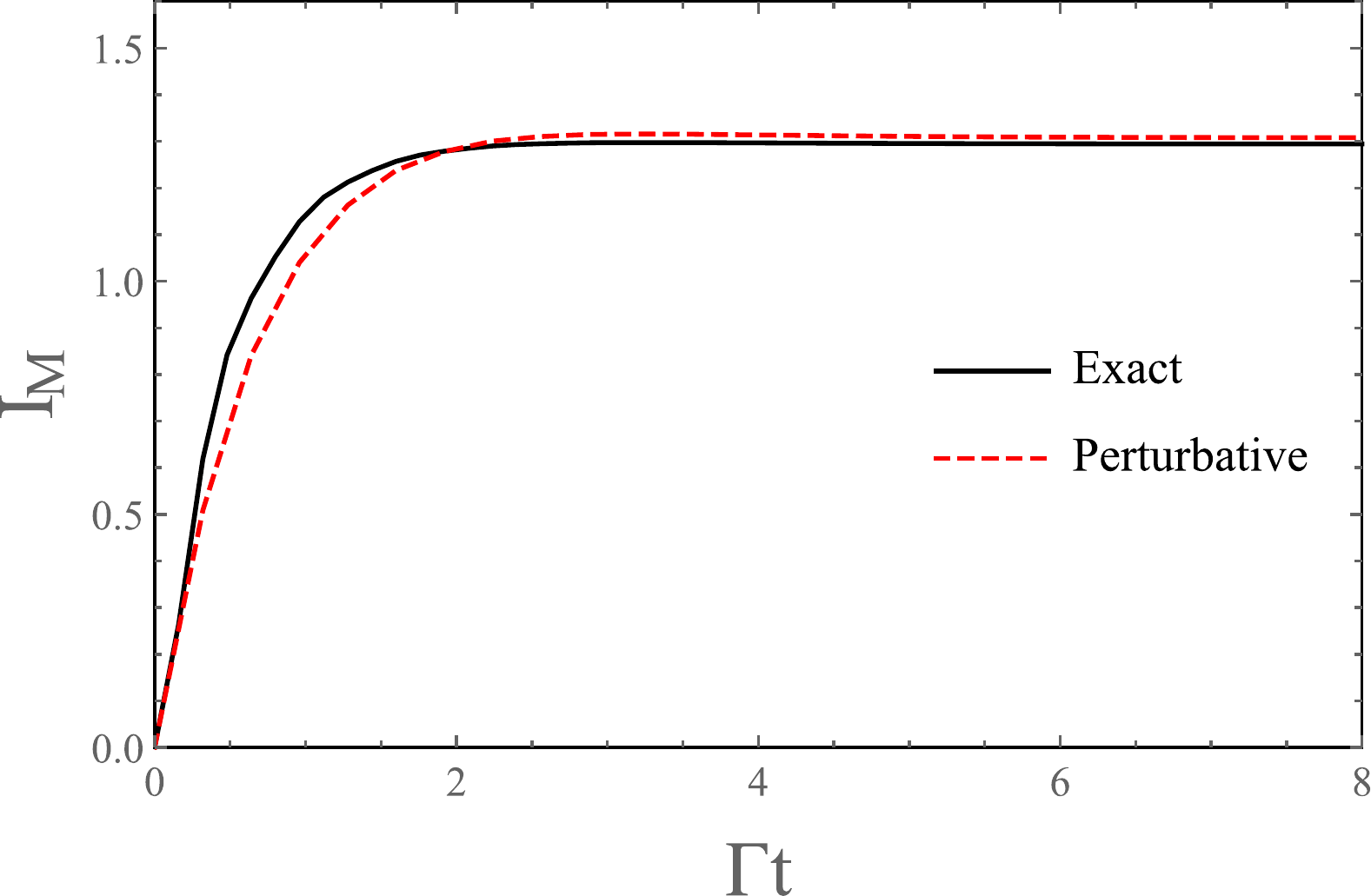}
	\caption{Exact total correlation (black solid line) compared with the perturbative formula~\eqref{pertitot} (red dashed line) as a function of time. Parameters as in Fig.~\ref{fig:cormatdyn}.}
	\label{fig:corsum400}
\end{figure}

\begin{figure}
	\centering
	\includegraphics[width=0.9\linewidth]{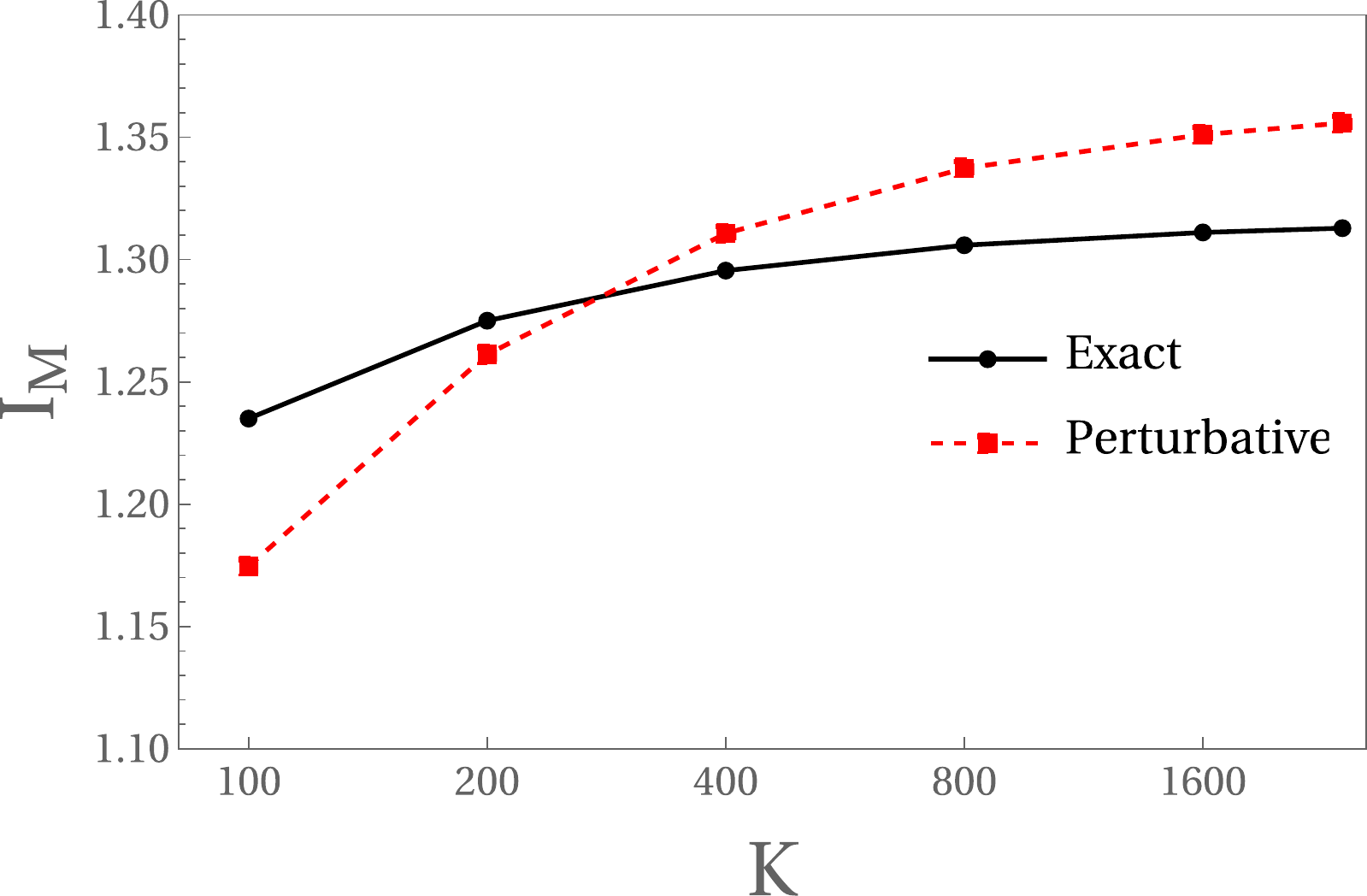}
	\caption{Scaling of the exact total correlation (black dots) compared with the perturbative formula~\eqref{pertitot} (red squares) as a function of the number of modes $K$ at a fixed time $\Gamma t=5$. Other parameters as in Fig.~\ref{fig:cormatdyn}. Lines shown for eye guidance.}
	\label{fig:corsum-scaling}
\end{figure}

Finally, let us check how appropriately the total correlation $I_M$ can be evaluated using the perturbative formula~\eqref{pertitot}. In Fig.~\ref{fig:corsum400} we present the evolution of the exact and approximate total correlation for $K=400$ modes in the bath, while Fig.~\ref{fig:corsum-scaling} shows their values at a fixed time for different sizes of the environment. As can be observed, the approximate value of $I_M$ is not equal, though very similar to the exact one. Furthermore, it does not converge to the exact result when one increases the bath size; rather, the perturbative formula tends to overestimate $I_M$ by a few percent (note the scale of the $y$ axis in Fig.~\ref{fig:corsum-scaling}). The origin of this discrepancy appears to be the presence of redundant correlations, due to which $I_M$ does not exactly correspond to the sum of two-point correlations $I_{ij}$. Nevertheless, though Eq.~\eqref{pertitot} is not exact even in the large bath limit, it still very well describes the order of magnitude of the total correlation, which supports the analytic argument presented in Sec.~\ref{sec:ferman}.

\subsubsection{Excursus: System-environment correlations} \label{sec:fermnumsecor}
While in our paper we focus on multipartite correlations $I_M$ and $J_M$, let us now take a small detour to consider quantum and classical contributions to bipartite correlations between the system and the environment. We note that the system-environment quantum mutual information $I_{SE}=S_S+S_E-S_{SE}$ is a part of the total correlation,
\begin{align} \label{totcordecomp}
	I_M=I_{SE}+I_\text{env},
\end{align}
where $I_\text{env}=\sum_{k=1}^K S_k-S_E$ is the intraenvironment correlation. To obtain the classical correlation $J_{SE}$ we apply a unitary Householder transformation~\cite{householder1958} acting on the environment which converts the correlation matrix to a form $\tilde{\mathcal{C}}$ in which all off-diagonal elements $\tilde{\mathcal{C}}_{0j}$ apart from $\tilde{\mathcal{C}}_{01}$ are equal to zero (see Ref.~\cite{ozaki} for a simple algorithm for non-Hermitian matrices). Then the classical system-environment correlation becomes fully concentrated in a pair of modes $0$ and $1$:
\begin{align}
	J_{SE}=J_{01}.
\end{align}
This two-mode correlation is then easily computable (see Sec.~\ref{sec:ferm2mode}). We note that a similar procedure of the correlation localization has been previously demonstrated for bosonic Gaussian states~\cite{adesso2004, serafini2005}.

\begin{figure}
	\centering
	\includegraphics[width=0.9\linewidth]{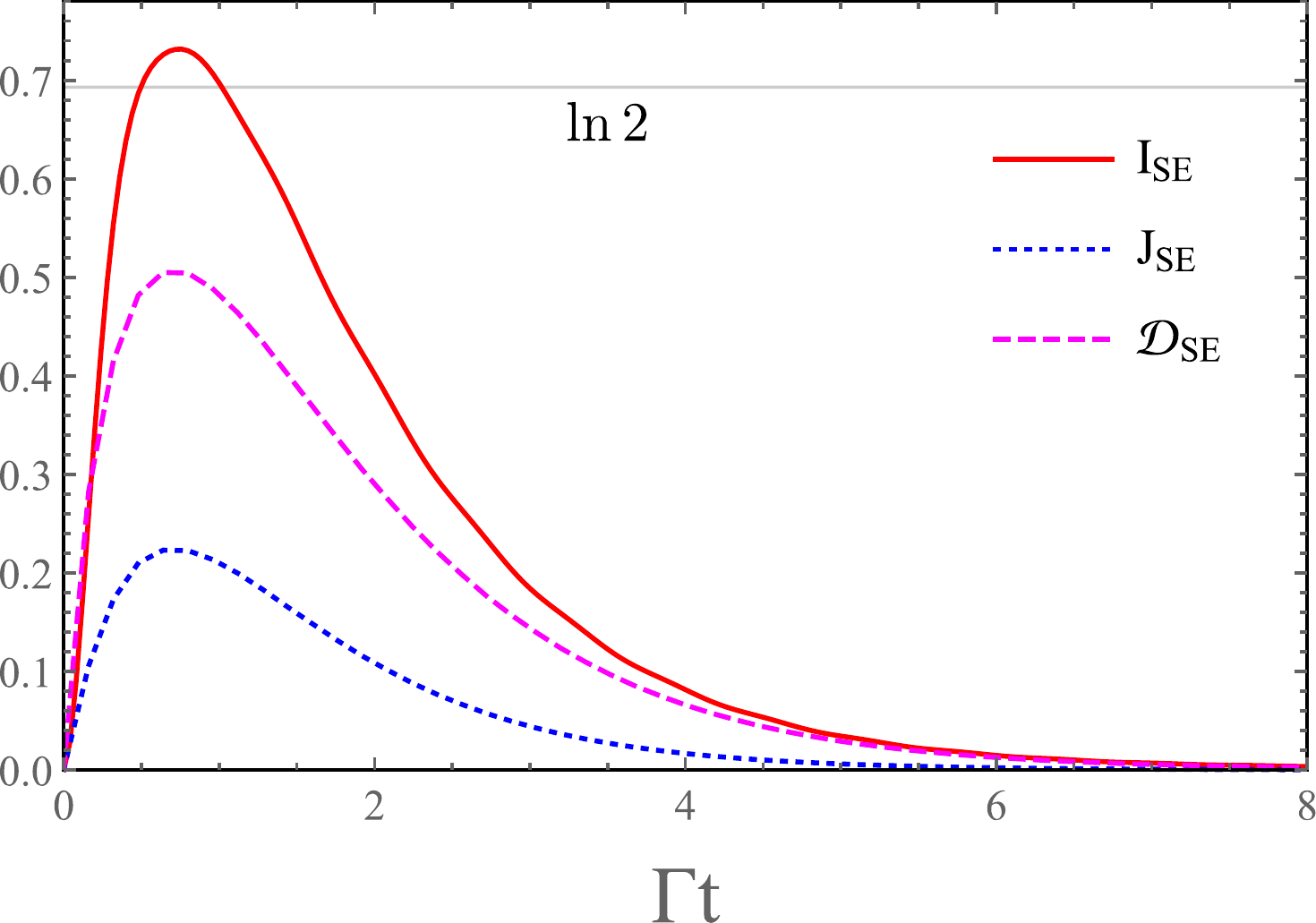}
	\caption{System-environment mutual information $I_{SE}$ (red solid line) compared with the classical correlation $ J_{SE}$ (blue dotted line) and the quantum discord $\mathcal{D}_{SE}=I_{SE}-J_{SE}$ (magenta dashed line). $\ln 2$ denotes a maximum bound for the mutual information in separable states. Parameters as in Fig.~\ref{fig:cormatdyn}.}
	\label{fig:cormatsecor}
\end{figure}
The results are presented in Fig.~\ref{fig:cormatsecor}. As one can observe, the mutual information between the system and the environment first increases but then tends to saturate at a low asymptotic value; this behavior is common for both fermionic~\cite{ptaszynski2022} and bosonic~\cite{pucci2013, colla2021, einsiedler2020} noninteracting systems. The term $J_{SE}$ is now non-negligible in comparison with the quantum mutual information $I_{SE}$ (especially at shorter timescales, when $I_{SE}$ is maximal). This can be explained using the equivalence of $J_{SE}$ to a correlation within a pair of modes: When $I_{SE}$ is large, the relative weight of the classical correlation increases (see Fig.~\ref{fig:ferm2s}); accordingly, when $I_{SE}$ decays to the asymptotic value, the weight of the classical contribution also decreases. Nevertheless, the quantum discord $\mathcal{D}_{SE}=I_{SE}-J_{SE}$ is still dominant on all timescales. Furthermore, we note that the quantum mutual information temporarily exceeds the value $\ln 2= \ln \text{dim}(\mathcal{H}_S)\geq S_S$ (where $\text{dim}(\mathcal{H}_S)$ is the dimension of the Hilbert space of the system), which is a maximum bound for the mutual information in separable (nonentangled) states~\cite{cerf1997, vollbrecht2002}. This implies the presence of (at least transient) system-environment entanglement, even though the reduced dynamics of the system is effectively classical and Markovian. We will explore this issue in a future study.

It may also be noted that the classical bipartite correlation $J_{SE}$ is actually larger than the classical multipartite correlation $J_M$ (cf.\@ Fig.~\ref{fig:cormatentr}). This might be surprising, as the total correlation obeys the inequality $I_M \geq I_{SE}$ [cf.\@ Eq.~\eqref{totcordecomp}]. The explanation is that $J_{SE}$ is maximized in a different measurement basis (obtained via the Householder transformation) than the base of modes diagonalizing the environment Hamiltonian (in which $J_M$ is calculated). From a technical point of view, this implies that the obtained value of $\text{max} J_M$ depends on the ordering of modes in the chain rule in Eqs.~\eqref{chain}--\eqref{holevo}; in particular, it would be overestimated if we wrote Eq.~\eqref{chain} in the form
\begin{align} 
	J_M \leq \sum_{k=0}^{K-1} J_{k+1 \ldots K,k},
\end{align}
since $\max J_{M}$ would then also include a contribution $J_{SE}$. On the other hand, this also suggests that one can actually access more information about correlations within the global system $SE$ than given by $J_M$ using the following procedure:
\begin{itemize}
	\item Partition of modes into $m \in [2,K+1]$ subsystems consisting of a single or several modes (e.g., for $m=2$, the system and the environment).
	\item Performing measurements on subsystems rather than individual modes.
	\item Calculation of the mutual information between the measurement outputs.
\end{itemize}
For purely classical correlations such coarse-graining would always lead to the information loss, as one loses track of correlations between modes belonging to a single subsystem; however, in quantum systems this can be compensated by gaining more freedom in choosing the measurement basis. Thus, in analogy to Eq.~\eqref{classcor}, the classical correlation can be alternatively defined as
\begin{align} \nonumber
	&\mathcal{J}_M=\max_{m,\alpha_m,\{\Pi_{\alpha_m,1}\},\ldots,\{\Pi_{\alpha_m,m}\}} \left[ \sum_{i=1}^m H(\mathcal{A}_{\alpha_m,i}) \right. \\
	&\left. -H(\mathcal{A}_{\alpha_m,1} \ldots \mathcal{A}_{\alpha_m,m}) \right],
\end{align}
where $\alpha_m$ denotes a single way of partitioning the global system $SE$ into $m$ subsystems and $H(\mathcal{A}_{\alpha_m,i})$ is the Shannon mutual information of local measurements on the subsystem $i$; then $J_M$ and $J_{SE}$ are special (unoptimized) instances of $\mathcal{J}_M$ for $m=K+1$ and $m=2$, respectively. This definition is inspired by the multipartite relative entropy of quantumness defined in Ref.~\cite{debarba2017}. However, since the calculation of this quantity is a complex combinatorial problem, its analysis goes beyond the scope of the present paper.

\subsection{Final remarks}
Let us now set our results in the context of literature. First, we note that -- as many-body Fock states are eigenstates of $\hat{H}_S+\hat{H}_E$ -- our result implies vanishing of the classical relative entropy between probability distributions of eigenstates of the environment since
\begin{align}
	D[\Pi_E(t)||\Pi_E(0)] \leq J_M+D_\text{env},
\end{align}
where $\Pi_E$ is the diagonal part of the density matrix of the environment in the eigenstate basis. The disappearance of such a relative entropy is tacitly implied in the derivation of the relation of heat and the observational entropy presented in Ref.~\cite{strasberg2021}, which is based on scaling arguments. However, since scaling arguments can sometimes lead to wrong predictions~\cite{ptaszynski2019}, an independent confirmation of this result is valuable.

Second, we note a similarity of our observations to the result of Kaszlikowski \textit{et al.}~\cite{kaszlikowski2008} showing that genuine multipartite quantum correlations can exist on their own without a supporting background of classical correlations. Although in Ref.~\cite{kaszlikowski2008} the classical correlations have been defined via correlations of observables rather than information-theoretic quantities, this observation can easily be generalized to the classical mutual information, as a generic probability distribution can be expressed by means of generalized moments [Eq.~\eqref{probfrommom}].

Finally, our observations may appear to contradict the previous results showing that the parity superselection rule reduces rather than increases the amount of quantum correlations~\cite{wiseman2003, ding2021, ernst2022}. However, there is no contradiction. In those previous studies, quantum correlations were defined as an \textit{accessible} entanglement, which can be used as a genuine quantum resource. The parity superselection rule reduces this accessibility via constrains on physically-allowed local operations. In contrast, in our case the correlations do not play the role of a useful resource, but rather a ``sink'' which allows the entropy production to grow even though the total von Neumann entropy $S_{SE}$ is conserved. This distinction between entanglement and discord has previously been stressed by Pusuluk \textit{et al.}~\cite{pusuluk2022}.
	
	\section{Bosonic systems} \label{sec:bos}
	\subsection{General discussion} \label{sec:bosan}
	
	We now turn our attention to noninteracting bosonic systems described by quadratic Hamiltonians
	\begin{align} \label{hambos}
		\hat{H}=\sum_{ij} \mathcal{A}_{ij} \hat{x}_i \hat{x}_j,
	\end{align}
where (for a generic $N$ mode system) $\hat{x}_{i}=\hat{q}_i$ ($i=1,\ldots,N)$ is the position operator and $\hat{x}_{i}=\hat{p}_{i-N}$ ($i={N+1},\ldots,2N)$ is the momentum operator; the position and momentum operators can be further expressed by means of bosonic creations and annihilation operators: $\hat{q}_i=(a_i^\dagger+a_i)/\sqrt{2}$ and $\hat{p}_i=i(a_i^\dagger-a_i)/\sqrt{2}$. As for fermionic systems, a thermal state of a quadratic Hamiltonian is a Gaussian state which can be wholly characterized by the $2N$ vector of the average moments $\mathbf{\bar{x}}=(\langle \hat{x}_1 \rangle,\ldots,\langle \hat{x}_{2N} \rangle)$ and the $2N \times 2N$ covariance matrix with the elements 
\begin{align}
	\Sigma_{ij}= \frac{1}{2} \langle \{\hat{x}_i,\hat{x}_j \} \rangle - \langle \hat{x}_i \rangle \langle \hat{x}_j \rangle.
\end{align}
The information-theoretic aspects of bosonic Gaussian states have been thoroughly investigated; see Refs.~\cite{ferraro2005, weedbrook2012, adesso2014} for comprehensive reviews. In particular, the von Neumann entropy of a Gaussian state can be calculated as~\cite{holevo1999}
\begin{align} \nonumber
	S=\sum_{i=1}^N &\left[ \left(\nu_i+\frac{1}{2} \right) \ln \left(\nu_i+\frac{1}{2} \right) \right. \\ &\left. -\left(\nu_i-\frac{1}{2} \right) \ln \left(\nu_i-\frac{1}{2} \right) \right],
\end{align}
where $\nu_i$ are the symplectic eigenvalues of $\Sigma$; they are equal to the positive eigenvalues of the matrix $i \Omega \Sigma$~\cite{ferraro2005, weedbrook2012, adesso2014} where
	\begin{align}
	\Omega	= \begin{pmatrix}
		0 & \mathds{1}_N \\
		-\mathds{1}_N & 0
	\end{pmatrix}
\end{align}
is the symplectic form, with $\mathds{1}_N$ being the $N \times N$ identity matrix.
	
	Analogously to the fermionic case, we will now express the covariance matrix of the system-environment ensemble as
	\begin{align}
		\Sigma=\Sigma^D+ \epsilon \Lambda,
	\end{align}
	where $\Sigma^D$ is the covariance matrix of uncorrelated modes and $\epsilon \Lambda$ describes the intermode correlations, with $\epsilon$ being a perturbation parameter. It can be now argued that, as for fermionic systems, to the lowest order of $\epsilon$ the mutual information in the Fock basis is a sum of pairwise mutual informations $J_{ij}^F$ and scales as $\mathcal{O}(\epsilon^4)$. This is because Wick's theorem~\eqref{wick} is applicable also to bosonic Gaussian systems [with $\sigma(P)$ equal to 1], and therefore the whole argumentation leading to Eq.~\eqref{entrfockexp2} is also valid. This might suggest that also in this case the intermode correlations are mostly quantum. However, there is a crucial difference between fermionic and bosonic systems: The parity superselection rule does not apply to bosonic systems, and thus the projective measurements on superpositions of Fock states are now allowed. For example, one of the most important measurements in the field of quantum optics -- the heterodyne measurement -- corresponds to a projection on the coherent state~\cite{yuen1980}
	\begin{align}
		| \alpha \rangle =e^{\frac{-|\alpha|^2}{2}} \sum_{n=0}^\infty \frac{\alpha^n}{\sqrt{n!}} |n \rangle,
	\end{align}
	which is a coherent superposition of different Fock states. Furthermore, coherences in the Fock basis are related to correlations between positions and momenta of different modes; for example, representing the density matrix of a pair of modes $i$ and $j$ as
	\begin{align}
		\rho = \sum_{n_i,n_j,m_i,m_j} \rho_{(n_i,n_j),(m_i,m_j)} |n_i,n_j \rangle \langle m_i,m_j |,
	\end{align}
	and expressing the creation operators as
	\begin{align}
		a_i^\dagger =\sum_{ni,nj} \sqrt{n_i+1} |n_i+1,n_j \rangle \langle n_i,n_j|,
	\end{align}
	one shows that correlations between positions of two modes are given by real parts of coherences in the Fock basis:
	\begin{align}
		&\langle \hat{q}_i \hat{q}_j \rangle =\frac{1}{2} \text{Tr} \left[ \rho (a_i^\dagger+a_i)(a_j^\dagger+a_j) \right]= \\ \nonumber
		&\sum_{n_i,n_j} \sqrt{(n_i+1) (n_j+1)} \times  \\ \nonumber &\left\{ \text{Re}[\rho_{(n_i,n_j),{(n_i+1,n_j+1)}}]+\text{Re}[\rho_{(n_i+1,n_j),{(n_i,n_j+1)}}] \right\}.
	\end{align}
	Quite obviously, such correlations may also be present in the classical, high temperature limit, where noninteracting bosonic systems correspond to networks of classical harmonic oscillators. As follows, in such a regime the entropy production should be purely classical in nature; we will confirm this numerically in Sec.~\ref{sec:bosnum}. This resembles the result of Li~\cite{li2019}, who related the entropy production in a gas of classical particles to the generation of position-momentum correlations. Furthermore, our observations are closely related to those of Smith \textit{et al.}~\cite{smith2022}, who noted a correspondence between the quantum coherence in the eigenstate basis and the inhomogenity of the probability density on a microcanonical shell in the classical phase space.
	
	To better estimate the amount of classical correlations we will therefore apply the Wehrl mutual information~\cite{floerchinger2021}
	\begin{align}
		J^W_M=S_S^W+\sum_{k} S_k^W-S_{SE}^W,
	\end{align}
	where $S^W$ is the Wehrl entropy~\cite{wehrl1978, wehrl1979}, which operationally corresponds to the Shannon entropy of the output of the heterodyne measurements~\cite{buzek1995}. Therefore, the Wehrl mutual information $J_M^W$ -- as the mutual information in the Fock basis $J^F_M$ -- is a classical mutual information between the measurement outputs, and thus [as implied by Eq.~\eqref{classcor}] provides a lower bound to the classical correlation $J_M$. The Wehrl entropy for the $N$--mode Gaussian state can be calculated as~\cite{floerchinger2021}
	\begin{align}
		S^W = \frac{1}{2} \ln \det \left(\Sigma+\frac{\mathds{1}_{2N}}{2} \right) + N.
	\end{align}
	
	\subsection{Two-mode correlations} \label{sec:bos2mode}
	We will now compare the magnitudes of the defined correlation measures considering a two-mode system with zero average moments ($\langle \hat{x}_i \rangle=0$). The covariance matrix of such a system can be transformed by local unitary transformations to a standard form~\cite{duan2000}
	\begin{align} \label{covmat2modbos}
		\Sigma= \begin{pmatrix} \langle n_i \rangle+\frac{1}{2} & \epsilon_q & 0 & 0 \\  \epsilon_q & \langle n_j \rangle+\frac{1}{2} & 0 & 0 \\ 
			0 & 0 & \langle n_i \rangle+\frac{1}{2} & \epsilon_p \\
			0 &0 & \epsilon_p & \langle n_j \rangle+\frac{1}{2} \end{pmatrix},
	\end{align}
	where $\langle n_i \rangle$ are the mean occupancies of the modes, and the parameters $\epsilon_q$ and $\epsilon_p$ are covariances or positions and momenta, respectively. In this section we focus on two-mode squeezed thermal states with $\epsilon_q=-\epsilon_p=\epsilon$~\cite{serafini2004}; as shown in the Appendix~\ref{sec:bos2modesym}, the main conclusions can be generalized to other symmetries of covariance matrix elements. The correlation terms are constrained as $\epsilon \leq \epsilon_\text{max}=\sqrt{(\langle n_i \rangle+1) \langle n_j \rangle}$ for $\langle n_i \rangle \geq \langle n_j \rangle$. The total mutual information $I_{ij}$ and the Wehrl mutual information $J_\text{ij}^W$ can be calculated analytically; expanding the obtained expressions into the power series of $\epsilon$ one gets
	\begin{align}
		I_{ij} &=\frac{\ln \left[ \frac{(1+\langle n_i \rangle) (1+\langle n_j \rangle)}{\langle n_i \rangle \langle n_j \rangle} \right]}{1+\langle n_i \rangle + \langle n_j \rangle } \epsilon^2 +\mathcal{O}(\epsilon^4), \\
		J_{ij}^W &=\frac{\epsilon^2}{ (1+\langle n_i \rangle)(1+\langle n_j \rangle)} +\mathcal{O}(\epsilon^4).
	\end{align}
	As can be observed, both $I_{ij}$ and $J_{ij}^W$ are of the order $\mathcal{O}(\epsilon^2)$. It may also be verified that both quantities coincide in the classical limit of large occupancy: $J^W_{ij} \rightarrow I_{ij}$ for $\langle n_i \rangle,{ }\langle n_j \rangle \rightarrow \infty$. Therefore, intermode correlations become purely classical for large occupancies. The classical correlation in the Fock basis has been calculated numerically using Eq.~\eqref{entrfock} with a finite cutoff $n_i,{ }n_j \leq 30$. The state probabilities $p(n_i,n_j)$ have been obtained using the Hermite polynomial approach developed by Dodonov \textit{et al.}~\cite{dodonov1994}; see Appendix~\ref{sec:fockprob} for details.
	
	\begin{figure}
		\centering
		\includegraphics[width=0.9\linewidth]{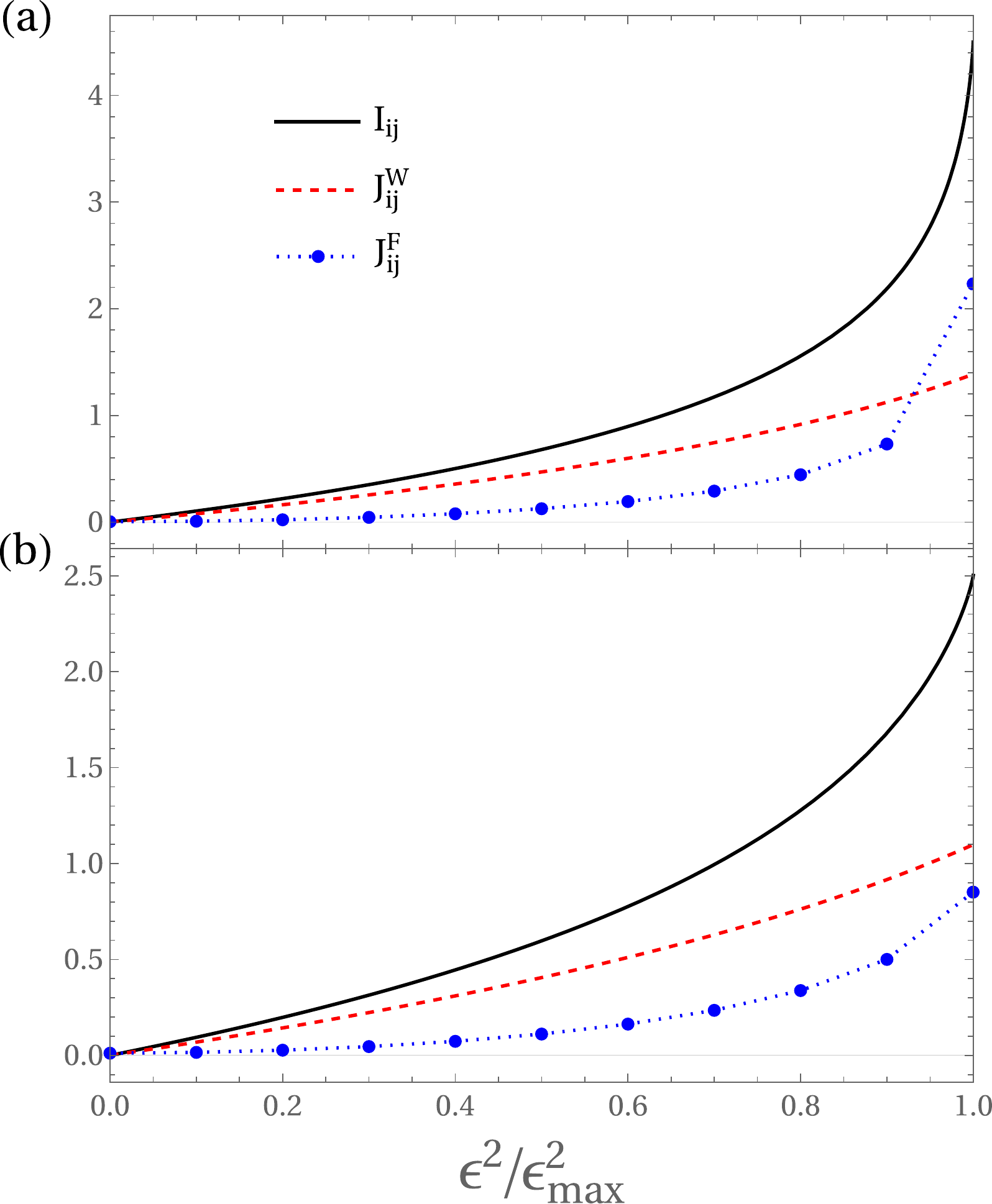}
		\caption{The total mutual information between two bosonic modes $I_{ij}$, the Wehrl mutual information $J^W_{ij}$, and the classical mutual information in the Fock basis $J_{ij}^F$ as a function of $\epsilon^2$ for (a) $\langle n_i \rangle=\langle n_j \rangle=3$, and (b) $\langle n_i \rangle=4$, $\langle n_j \rangle=2$, with $\epsilon_\text{max}=\sqrt{(\langle n_i \rangle+1)\langle n_j \rangle}$.}
		\label{fig:bos2s}
	\end{figure}
	
The results for two choices of mean occupancies, $\langle n_i \rangle=\langle n_j \rangle=3$ and $\langle n_i \rangle=4$, $\langle n_j \rangle=2$, are illustrated in Fig.~\ref{fig:bos2s}. As may be observed, for small $\epsilon$ the terms $I_{ij}$ and $J_{ij}^W$ are of a similar magnitude; therefore, correlations in the position-momentum phase space are predominantly classical. On the other hand, the classical correlation in the Fock basis $J^F_{ij}$ scales as $\epsilon^4$ instead of $\epsilon^2$, and is thus negligible for a small $\epsilon$. This confirms that for weakly correlated bosonic Gaussian states the Wehrl mutual information better estimates the magnitude of classical correlations than the classical mutual information in the Fock basis. 
	
Additionally, one may observe that the symmetry of covariance matrix elements plays a role in the high correlation regime $\epsilon \approx \epsilon_\text{max}$. Then for equal mean occupancies $\langle n_i \rangle=\langle n_j \rangle=3$ the state of the two mode-system becomes an entangled pure state $|\psi \rangle=\sum_{l=0}^\infty \alpha_l |l,l \rangle$ in the limit of $\epsilon = \epsilon_\text{max}$~\cite{schumaker1985}, and the mutual information in the Fock basis is equal to the half of quantum mutual information: $J_{ij}^F=I_{ij}/2$. In this case, $J_{ij}^F$ becomes dominant over $J_{ij}^W$ for large $\epsilon$. For unequal occupancies $\langle n_i \rangle=4$, $\langle n_j \rangle=2$ the state is mixed, and the Wehrl mutual information remains dominant also in the high correlation regime.
	
	\subsection{Numerical results} \label{sec:bosnum}
	The magnitude of the classical contribution to the entropy production is now further investigated by performing simulations of the time evolution of the covariance matrix for the Caldeira-Leggett model~\cite{caldeira1983} described by the Hamiltonian
	\begin{align} \label{hamcald}
		\hat{H}_{SE}=\sum_{i=0}^K \frac{\omega_i}{2} (\hat{q}_i^2+\hat{p}_i^2)- \sum_{i=1}^K \kappa_i \hat{q}_0 \hat{q}_i,
	\end{align}
where (as for the fermionic case) the index $i=0$ corresponds to the system and $i=1,\ldots,K$ to the modes of the environment. Details of the simulation approach are presented in the Appendix~\ref{sec:covev}. The energy levels of the bath are chosen as $\omega_j=j\omega_c/K$, where $\omega_c$ is the cutoff frequency. The couplings $\kappa_i$ are parameterized as $\kappa_i=\sqrt{2 J(\omega) \Delta \omega/\pi}$, where $\Delta \omega=\omega_c/K$ and $J(\omega)=\gamma \omega$ is the Ohmic spectral density. The system is initialized in the state with a null vector of average moments ($\mathbf{\bar{x}}=0$) and the covariance matrix
\begin{align}
	\Sigma(0)=\text{diag}(\varsigma,\varsigma)+\frac{1}{2} \mathds{1}_{2K+2},
\end{align}
where $\varsigma={\text{diag}[\langle \hat{N}_0(0) \rangle,n(\omega_1),\ldots,n(\omega_K)]}$ and $n(\omega)={[\exp(\beta \omega)-1]^{-1}}$ is the Bose-Einstein distribution. The energy of the environment is calculated as
\begin{align}
	\langle \hat{H}_E \rangle=\sum_{i=1}^K \frac{\omega_i}{2} \left(\Sigma_{ii}+\Sigma_{K+1+i,K+1+i} \right),
\end{align}
since $\Sigma_{ii}=\langle \hat{q}_i^2 \rangle$ and $\Sigma_{K+1+i,K+1+i}=\langle \hat{p}_i^2 \rangle$.

	\begin{figure}
		\centering
		\includegraphics[width=0.9\linewidth]{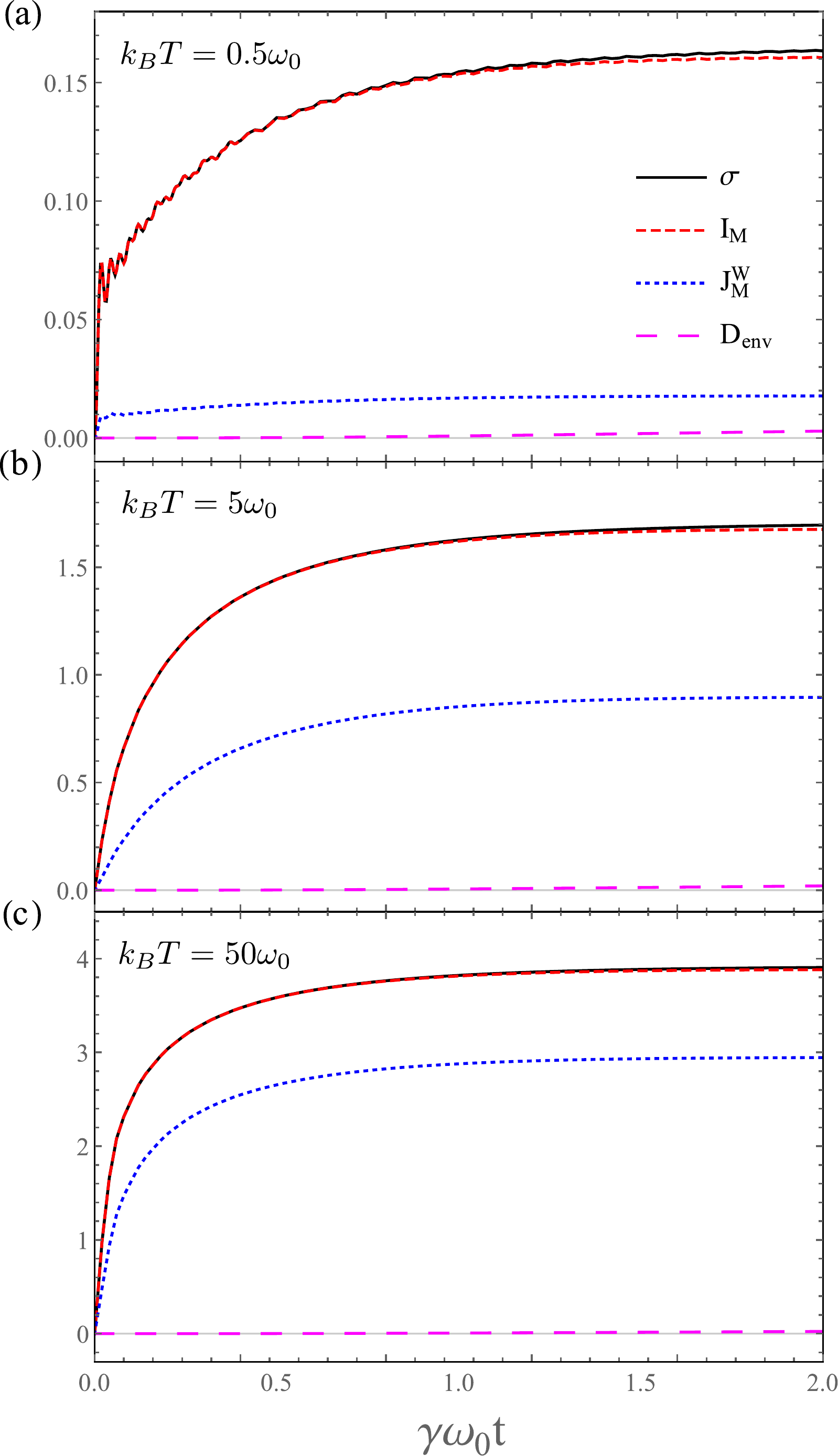}
		\caption{Entropy production and its constituents for the Caldeira-Leggett model with the initial vacuum state of the system, $\omega_c=4\omega_0$, $\gamma=0.01 \omega_0$, and $K=600$.}
		\label{fig:bosvac}
	\end{figure}
	
	\begin{figure}
		\centering
		\includegraphics[width=0.9\linewidth]{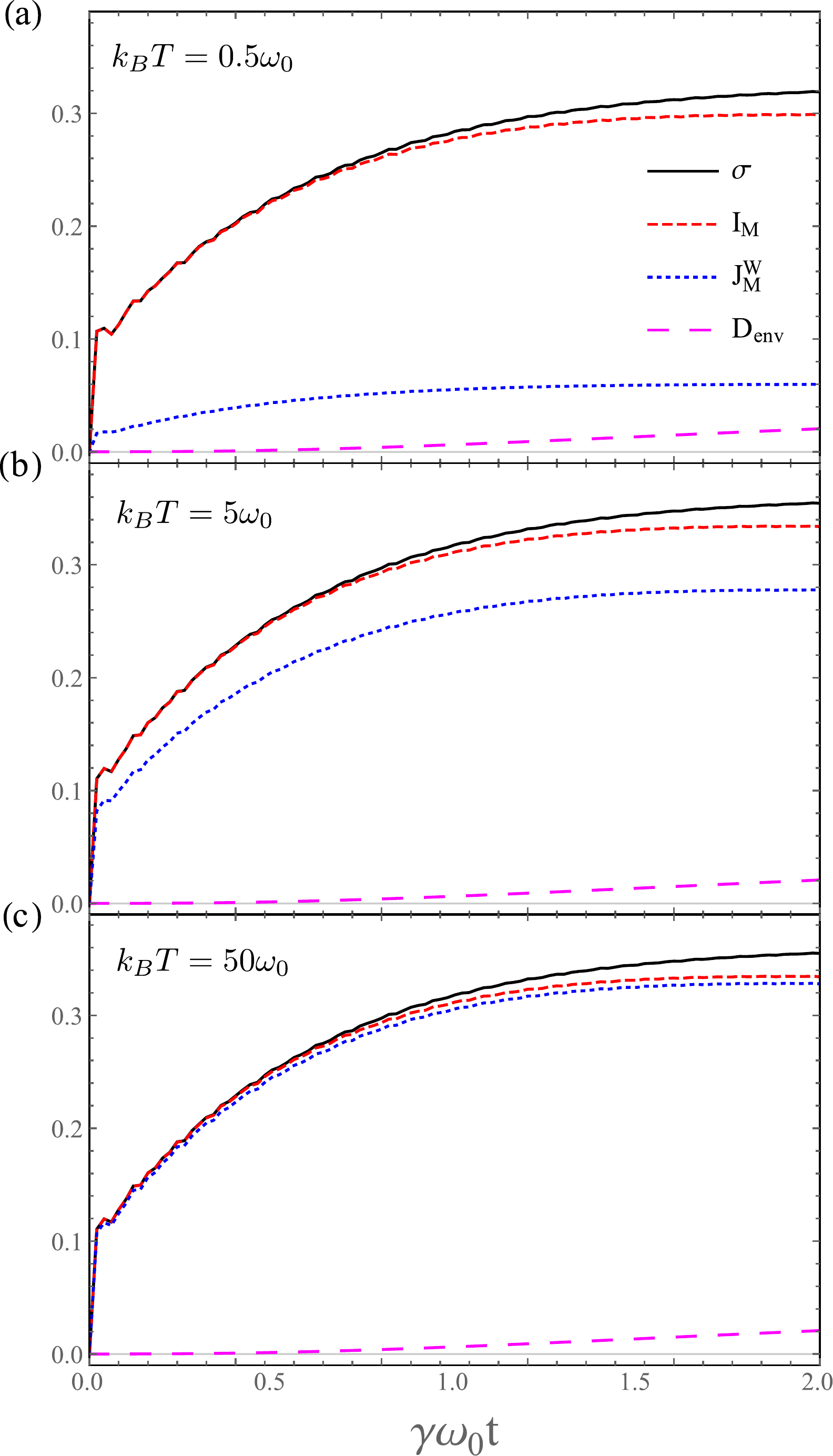}
		\caption{Entropy production and its constituents for the Caldeira-Leggett model with the system initialized in a thermal state with the temperature $2T$ and other parameters as in Fig.~\ref{fig:bosvac}.}
		\label{fig:bosth}
	\end{figure}
	
	We analyze two cases. First, as in Refs.~\cite{pucci2013, einsiedler2020, colla2021}, we consider thermalization of a system initialized in the ground (vacuum) state for different temperatures of the bath $T$; results are presented in Fig.~\ref{fig:bosvac}. In the second case, presented in Fig.~\ref{fig:bosth}, the system is initialized in a thermal state with temperature $2T$. As one may observe, for both cases the Wehrl mutual information $J^W_M$ is relatively small for low temperatures, while it becomes a dominant contribution to the entropy production in the high temperature limit. Therefore, one may conclude that the correlations responsible for the entropy production undergo a quantum-to-classical transition when the temperature increases. However, the magnitude of quantum correlations depends not only on the temperature of the bath but also on the initial state of the system. For the initial vacuum state, the Wehrl mutual information does not coincide with the total correlation even for high temperatures of the bath [Fig.~\ref{fig:bosvac}~(c)], while this does take place for the initial high temperature thermal state [Fig.~\ref{fig:bosth}~(c)]. This may be explained as follows: When both the system and the environment are initialized in the effectively classical high temperature thermal states, the generated correlations are fully classical. On the other hand, when the system is initialized in the high-purity vacuum state and coupled to a high temperature environment, the state of the system becomes classical due to thermalization, while its initial ``quantumness'' (here related just to impossibility of joint measurement of position and momentum) is reconverted into the quantum correlation between the modes. This resembles the result of Ref.~\cite{eisert2002}, showing that when the system is initialized in a pure state, there is always a generation of entanglement between the system and the environment, regardless of temperature and the system-bath coupling. Furthermore, one may relate our result to a standard interpretation of decoherence as a reconversion of the initially local quantum information of the system into quantum correlations with the environment~\cite{zurek2003}.
	
We note that the apparent lack of reconversion of the system quantumness into classical correlations may be related to the quadratic nature of the considered Hamiltonian; in such a case the same equations of motions for the covariance matrix apply to both quantum and classical models~\cite{ullersma1966}, and thus the dynamics of quantum and classical correlations is decoupled. The open issue is whether different outcomes may be provided by non-quadratic Hamiltonians, e.g., the spin-boson model; however, this goes beyond the scope of the present paper.
	
	\section{Semiclassical transport in the low density limit} \label{sec:lowden}
	So far, we have treated fermionic and bosonic systems in a separate way, showing their different behavior with respect to the nature of microscopic correlations responsible for the entropy production. There is, however, a limit in which the properties of fermionic and bosonic particles can be expected to converge: the low density limit where the mode occupancies $\langle \hat{N}_i \rangle$ are close to zero. It is reached when the level energies (with respect to the chemical potential) are much larger than the thermal energy $k_B T$, such that the Fermi and Bose-Einstein distributions converge to the Maxwell-Boltzmann distribution, i.e.,
	\begin{align}
		f(\omega) \approx n(\omega) \approx e^{-\beta \omega} \ll 1 \quad \text{for} \quad \omega \gg k_B T,
	\end{align}
	where we took $\mu=0$. In this regime, quantum correlations (such as the Pauli exclusion principle for fermions or particle bunching for bosons) play a negligible role, since the probability of ``meeting'' of two particles is small, and thus the properties of both fermions and bosons are expected to converge to those of classical particles~\cite{cheng2006}. This raises the question whether the microscopic correlations responsible for the entropy production also become classical in this limit.

To deal with this problem, we will analyze the microscopic constituents of the entropy production for the case of heat transport between two baths with different temperatures rather than relaxation of the system attached to a single bath (as in Secs.~\ref{sec:ferm}--\ref{sec:bos}). This is because in such a transport scenario the physical manifestation of the low density limit becomes most clear: The Levitov-Lesovik formula~\cite{levitov1993, saito2007, gaspard2015}, which describes the fluctuations of fermionic and bosonic currents, converges to an analogous equation derived for classical ballistic particles obeying the Maxwell-Boltzmann distribution~\cite{brandner2018}. In other words, the transport becomes effectively classical in terms of macroscopic observable quantities, such as heat currents. However, as we will show in Secs.~\ref{sec:lowdenferm}--\ref{sec:lowdenbos}, for both fermionic and bosonic cases the entropy production is then dominated by quantum intermode correlations. Therefore, there is no direct connection between the possibility of a semiclassical description of macroscopic transport quantities and the nature of microscopic correlations.
 	
	\subsection{Low density limit in nonequilibrium transport} \label{sec:levles}
	To illustrate the physics of the low density limit, let us first show how it manifests itself in the case of nonequilibrium heat transport between two baths. In noninteracting systems, fluctuation of both fermionic and bosonic currents can be described by the Levitov-Lesovik formula for the scaled cumulant generating function~\cite{levitov1993, saito2007, gaspard2015}. For a heat transport from the hot bath $H$ to the cold bath $C$ with a same chemical potential $\mu_H=\mu_R=0$ it takes the form
	\begin{align} \label{levitov} 
		&	\chi(\lambda) =
		\pm \int_{\omega_\text{MIN}}^{\omega_\text{MAX}} \frac{d \omega}{2\pi} \ln \left \{1 \pm \mathcal{T}(\omega) \times \right. \\ \nonumber
		& \left. \left [\left (e^{\lambda \omega}-1 \right)   g^{\pm}_H (\omega) h^{\pm}_C (\omega)  +\left (e^{-\lambda \omega}-1 \right) g^{\pm}_C (\omega) h^{\pm}_H (\omega) \right] \right \},
	\end{align}
where the sign $+$ ($-$) corresponds to fermions (bosons), $\mathcal{T}(\omega)$ is the transmission function, $g_\alpha^{\pm}(\omega) ={[\exp(\beta_\alpha \omega) \pm 1]}^{-1}$ is either the Fermi (+) or Bose-Einstein (-) distribution, $h_\alpha^{\pm}(\omega)=1 \mp g_\alpha^{\pm}(\omega)$, and $[\omega_\text{MIN},\omega_\text{MAX}]$ is the transport window, that is, the energy region with a finite transmission function. By assuming that $\omega \gg k_B T$ in the whole transport window $[\omega_\text{MIN},\omega_\text{MAX}]$, the formula simplifies to the same form for both fermions and bosons:
	\begin{align} \label{levitovbal} 
	\chi(\lambda) =\int_{\omega_\text{MIN}}^{\omega_\text{MAX}} \frac{d \omega}{2\pi} \mathcal{T}(\omega) \left[e^{-(\beta_H-\lambda)\omega}+e^{-(\beta_C+\lambda)\omega} \right]+C,
\end{align}
where $C$ is a constant. An equivalent formula has been derived by Brandner \textit{et al.}~\cite{brandner2018} for classical ballistic particles obeying the Maxwell-Boltzmann distribution. Therefore, in the low density limit quantum correlations cease to play a role in the transport, which becomes effectively classical.

\begin{figure}
	\centering
	\includegraphics[width=0.9\linewidth]{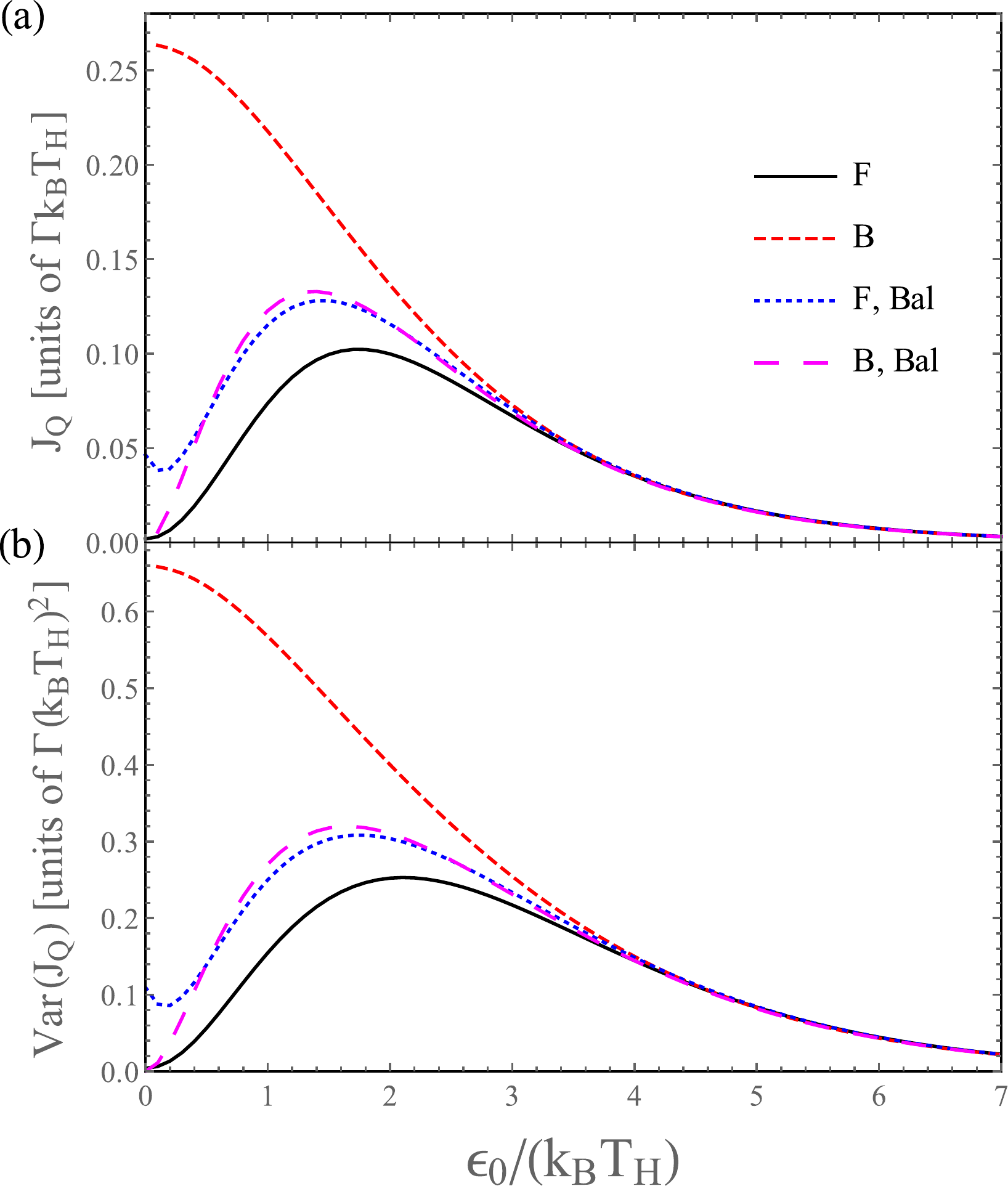}
	\caption{(a) The average heat current and (b) the current variance for fermions (F) and bosons (B) compared with values calculated using formula~\eqref{levitovbal} derived for classical ballistic particles with either fermionic or bosonic transmission function (F,Bal and B,Bal, respectively). Parameters: $T_C=0.5T_H$, $\Gamma=0.05k_B T_H$, $W=k_B T_H$, $\omega_c=3 \omega_0$}
	\label{fig:levitov-lesovik}
\end{figure}
To show that explicitly, we will analyze the average heat current and the current variance
\begin{align} \label{heatcur}
	J_Q &= \left[\frac{\partial}{\partial \lambda} \chi(\lambda) \right]_{\lambda=0}, \\
	\text{Var}(J_Q)&=\left[\frac{\partial^2}{\partial^2 \lambda} \chi(\lambda) \right]_{\lambda=0}.
\end{align}
For fermions, we will take a transmission function of a single energy level with energy $\epsilon_0$~\cite{haug2008}:
\begin{align} \label{transferm}
	T(\omega)=\frac{\Gamma_H \Gamma_C}{(\omega-\epsilon_0)^2+(\Gamma_H+\Gamma_C)^2/4},
\end{align}
where $\Gamma_\alpha$ is the coupling strength to the bath $\alpha$ and the transport window reads $[\omega_\text{MIN},\omega_\text{MAX}]=[\epsilon_0-W/2,\epsilon_0+W/2]$. An analogous formula for a single bosonic level with energy $\omega_0$ reads~\cite{agarwalla2017}
\begin{align}
	T(\omega)=\frac{4 \omega^2 J_H(\omega)J_C (\omega)}{(\omega^2-\omega_0^2)^2 +\omega^2 [J_H(\omega)+J_C(\omega)]^2},
\end{align}
where $J_\alpha(\omega)=\gamma_\alpha \omega$ is the Ohmic spectral density and $[\omega_\text{MIN},\omega_\text{MAX}]=[0,\omega_c]$. To allow for a comparison, we further take $\omega_0=\epsilon_0$ and $\Gamma_H=\Gamma_C=\gamma_H \omega_0=\gamma_C \omega_0=\Gamma$.

The results are presented in Fig.~\ref{fig:levitov-lesovik}. As one can observe, the transport statistics of bosons and fermions inedeed coincide and converge to that of classical ballistic particles for large level energies $\epsilon_0 \gtrapprox 4 k_B T$. This shows that the transport in the low density regime becomes effectively classical at the level of current statistics.

Additionally, we note that the transport properties of fermions and bosons differ drastically in the opposite limit of low level energies. In particular, in the limit $\epsilon_0 \rightarrow 0$ the heat current vanishes for fermions, while it is finite (and, in fact, maximal) for bosons. This can be explained as follows: For a weak coupling $\Gamma \ll k_B T_C$ the heat current can be approximated as
\begin{align}
    J_Q=\epsilon_0 \frac{\Gamma_H \Gamma_C} {\Gamma_H+\Gamma_C} \left[g^\pm_H(\epsilon_0) - g_C^\pm(\epsilon_0) \right].
\end{align}
For fermions $g_\alpha^+(\epsilon_0) \rightarrow 1/2$ in the particle-hole symmetric case of $\epsilon_0 \rightarrow 0$ (independent of the temperature), and thus the current vanishes. In contrast, for bosons $g_\alpha^-(\epsilon_0) \rightarrow k_B T_\alpha/\epsilon_0$ for $\epsilon_0 \rightarrow 0$, and thus the current takes a finite value $J_Q \rightarrow k_B (T_H-T_C)\Gamma_H \Gamma_C/(\Gamma_H+\Gamma_C)$ (in Fig.~\ref{fig:levitov-lesovik} it is a little higher due to the finite coupling effects).

\subsection{Fermionic case} \label{sec:lowdenferm}
We will now analyze the constituents of the entropy production in the low density limit, starting from the fermionic case. We consider a two-bath generalization of the Hamiltonian~\eqref{hamnrl}:
\begin{align} \nonumber
	\hat{H}_{SE}=&\epsilon_0 c^\dagger_0 c_0 +\sum_{\alpha=H,C} \sum_{k=1}^K \epsilon_{\alpha i} c_{\alpha i}^\dagger c_{\alpha i} \\ &+ \sum_{\alpha=H,C} \sum_{k=1}^K \left( t_{\alpha i} c^\dagger_0 c_{\alpha i} + \text{h.c.} \right).
\end{align}
As before, the bath levels are uniformly distributed over the interval $[\epsilon_0-W/2,\epsilon_0+W/2]$ and the tunnel couplings are parameterized as $t_{\alpha i}=\sqrt{\Gamma_\alpha W/[2\pi(K-1)]}$. The system is initialized in the state with a stationary occupancy
\begin{align}
\langle \hat{N}_0^\text{st} \rangle=\frac{\Gamma_H f_H(\epsilon_0)+\Gamma_C f_C(\epsilon_0)}{\Gamma_H+\Gamma_C},
\end{align}
where $f_\alpha(\epsilon)={\{1+\exp[\beta_\alpha(\epsilon-\mu_\alpha)]\}^{-1}}$; later we take $\Gamma_H=\Gamma_C=\Gamma$ and $\mu_H=\mu_C=0$ for simplicity. Accordingly, the initial correlation matrix is defined as
\begin{align} \nonumber
\mathcal{C}(0)=\text{diag}[&\langle \hat{N}_0^\text{st} \rangle,f_H(\epsilon_{H1}),\ldots,f_H(\epsilon_{HK}), \\ &f_C(\epsilon_{C1}),\ldots,f_C(\epsilon_{CK})].
\end{align}
The entropy production is compared with a prediction of the Levitov-Lesovik formula for classical ballistic particles~\eqref{levitovbal}
\begin{align}
	\sigma =(\beta_C-\beta_H) J_Q t,
\end{align}
where the heat current $J_Q$ is given by Eq.~\eqref{heatcur} and the fermionic transmission function~\eqref{transferm} is used. Furthermore, we also make a comparison with the results provided by the Markovian master equation
\begin{align}
	\sigma = \epsilon_0 (\beta_C-\beta_H) \frac{\Gamma_H \Gamma_C t}{\Gamma_H+\Gamma_C} \left[f_H(\epsilon_0) - f_C(\epsilon_0) \right].
\end{align}

\begin{figure}
	\centering
	\includegraphics[width=0.9\linewidth]{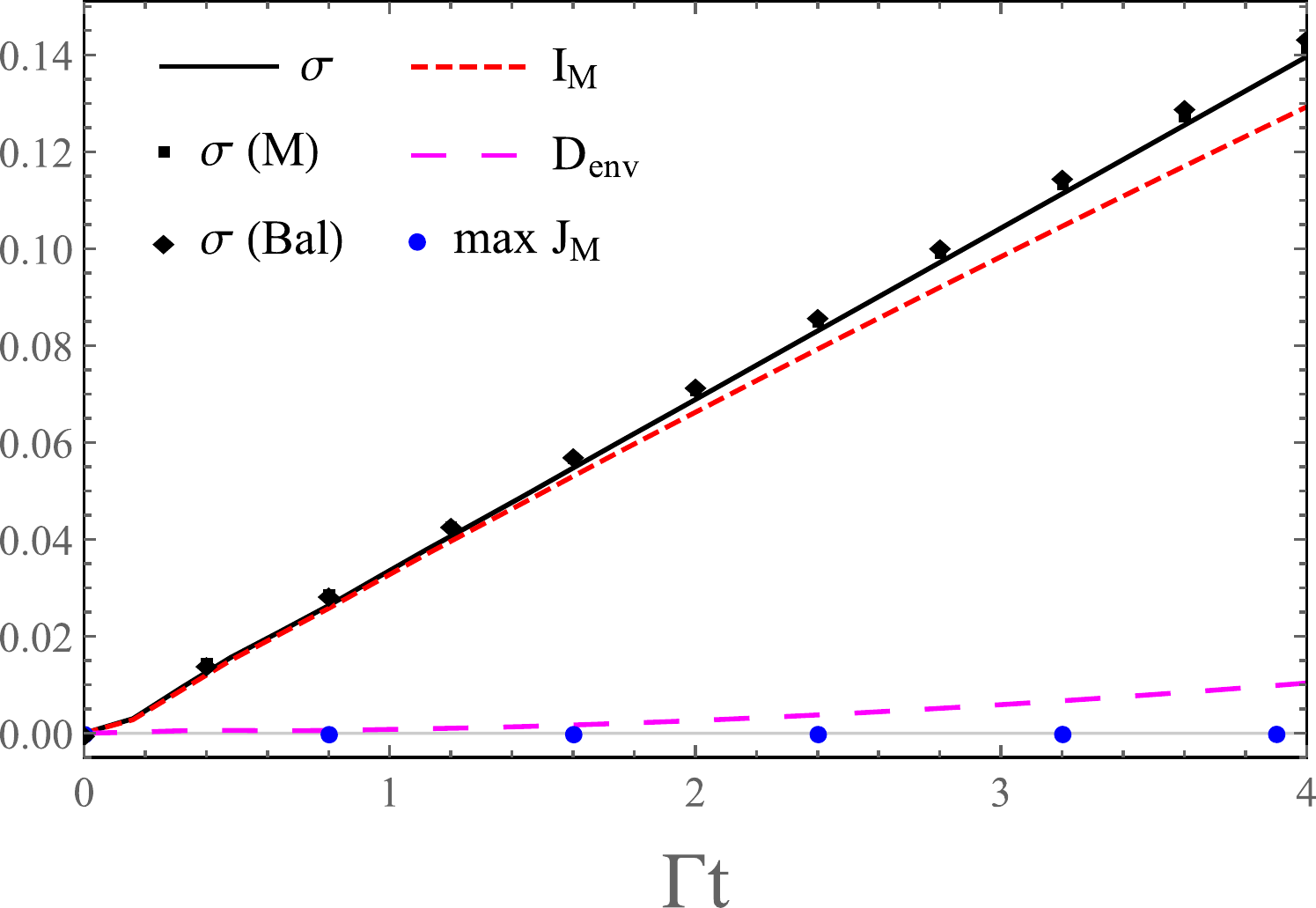}
	\caption{The entropy production and its constituents as a function of time for an initial stationary state; results compared with the prediction of the Markovian master equation (M) and formula~\eqref{levitovbal} derived for classical ballistic particles (Bal). Parameters: $\epsilon_0=4k_B T_H$, $\mu_H=\mu_C=0$, $T_C=0.5T_H$, $\Gamma_H=\Gamma_C=\Gamma=0.05k_B T_H$, $W=k_B T_H$, $K=400$.}
	\label{fig:ferm-lowden}
\end{figure}

The results are presented in Fig.~\ref{fig:ferm-lowden}. As may be observed, the entropy production agrees well with the predictions of the classical transport formula~\eqref{levitovbal} and the Markovian master equation. At the same time, the classical intermode correlation $J_M$ is completely negligible; in fact, the obtained upper bound is even lower than in the high-density case presented in Fig.~\ref{fig:cormatentr}. This result can be explained by analyzing the correlations within a two-mode system with a correlation matrix given by Eq.~\eqref{cormat2modferm}. Taking $\langle n_i \rangle = \langle n_j \rangle = \langle n \rangle \ll 1$ one obtains approximate formulas
\begin{align}
	I_{ij} & \approx \langle n \rangle \ln \frac{\langle n \rangle^2-\epsilon^2}{\langle n \rangle^2} + \epsilon \ln \frac{\langle n \rangle+\epsilon}{\langle n \rangle-\epsilon}, \\
	J_{ij}^F &\approx \epsilon^2 + (\langle n \rangle^2-\epsilon^2) \ln \frac{\langle n \rangle^2-\epsilon^2}{\langle n \rangle^2}.
\end{align}
As can be seen, the total correlation is proportional to the average occupancy $\langle n \rangle$, while the classical correlation to $\langle n \rangle^2$; therefore, the classical term becomes negligible for low occupancy. On the other hand, one notes that the contribution $D_\text{env}$ is finite and increases at larger times; this is because for finite baths a relative perturbation of mode occupancies is more significant in the low density limit.


\subsection{Bosonic case} \label{sec:lowdenbos}
Let us now turn our attention to bosons. We will consider a two-bath generalization of the Caldeira-Leggett model analyzed in Sec.~\ref{sec:bosnum}:
\begin{align} \nonumber
	\hat{H}_{SE}=&\frac{\omega_0}{2} (\hat{q}_0^2+\hat{p}_0^2)+\sum_{\alpha=H,C} \sum_{i=1}^K \frac{\omega_{\alpha i}}{2} (\hat{q}_{\alpha i}^2+\hat{p}_{\alpha i}^2)
	\\ &- \sum_{\alpha=H,C} \sum_{i=1}^K \kappa_{\alpha i} \hat{q}_0 \hat{q}_{\alpha i}.
\end{align}
We found a best convergence with the Markovian evolution by choosing the energy levels in the bath to be uniformly distributed throughout the interval $[\omega_0-W/2,\omega_0+W/2]$ (quite nonstandard parameterization compared to previous studies~\cite{colla2021, pucci2013, einsiedler2020}). The couplings $\kappa_{\alpha i}$ are parameterized as $\kappa_{\alpha i}=\sqrt{2 J_\alpha(\omega) \Delta \omega/\pi}$, where $\Delta \omega=W/(K-1)$ and $J_\alpha(\omega)=\gamma_\alpha \omega$. The system is initialized in the state with a stationary occupancy
\begin{align}
	\langle \hat{N}_0^\text{st} \rangle=\frac{\gamma_H n_H(\omega_0)+\gamma_C n_C(\omega_0)}{\gamma_H+\gamma_C},
\end{align}
where $n_\alpha(\omega)={[1+\exp(\beta_\alpha \omega)]^{-1}}$; we later take $\gamma_H=\gamma_C=\gamma$. As in the fermionic case, the entropy production is compared with the predictions of the Levitov-Lesovik formula for classical ballistic particles~\eqref{levitovbal}, as well as with the classical Markovian master equation
\begin{align}
	\sigma = \omega_0^2 (\beta_C-\beta_H) \frac{\gamma_H \gamma_C t}{\gamma_H+\gamma_C} \left[n_H(\omega_0) - n_C(\omega_0) \right].
\end{align}

\begin{figure}
	\centering
	\includegraphics[width=0.9\linewidth]{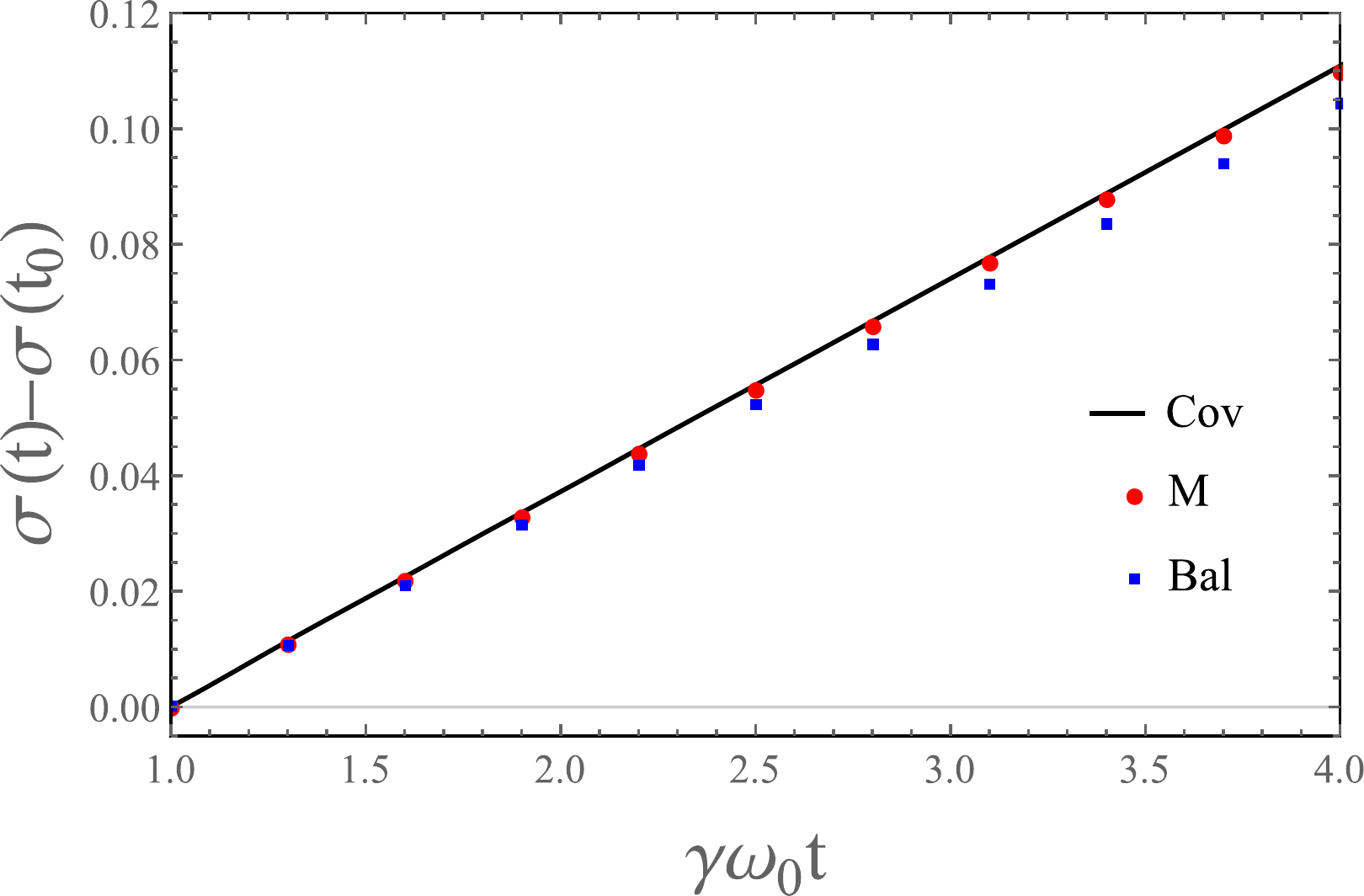}
	\caption{The stationary entropy production as a function of time for the covariance matrix approach (Cov), Markovian master equation (M), and formula~\eqref{levitovbal} derived for classical ballistic particles (Bal). Parameters: $t_0=1/(\omega_0 \gamma)$, $\omega_0=4k_B T_H$, $T_C=0.5T_H$, $\gamma_H=\gamma_C=\gamma=0.0125$, $W=k_B T_H$, $K=300$.}
	\label{fig:bos-lowden-dyn}
\end{figure}

\begin{figure}
	\centering
	\includegraphics[width=0.9\linewidth]{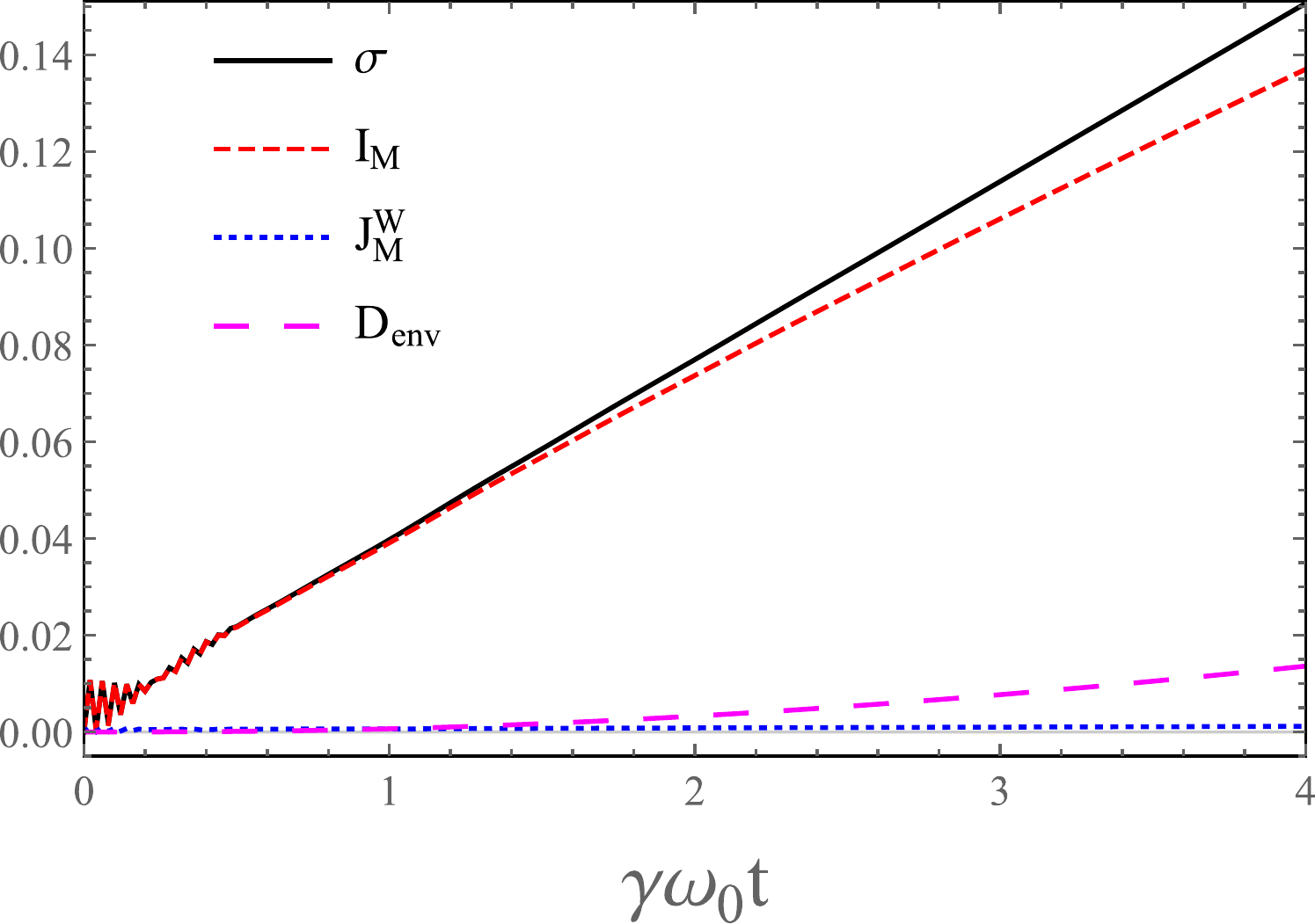}
	\caption{The entropy production and its constituents as a function of time for an initial stationary state. Parameters as in Fig.~\ref{fig:bos-lowden-dyn}.}
	\label{fig:bos-lowden-entr}
\end{figure}

The results are presented in Figs.~\ref{fig:bos-lowden-dyn}--\ref{fig:bos-lowden-entr}. One can note that at short times the entropy production exhibits non-Markovian oscillating dynamics. However, a good agreement with the master equation, and little worse with the ballistic formula, is observed in the stationary phase by analyzing a difference $\sigma(t)-\sigma(t_0)$, where $t_0$ is greater than the duration of the transient phase [Fig.~\ref{fig:bos-lowden-dyn}]. As shown in Fig.~\ref{fig:bos-lowden-entr}, the entropy production is dominated by the total intermode correlation, while the effectively classical Wehrl mutual information $J_M^W$ is negligible. This is, in fact, not surprising, since the low density limit corresponds to low temperatures. As in the fermionic system, the scaling of these two quantities can be explained by analyzing a two-mode system described by the covariance matrix~\eqref{covmat2modbos}. In the limit of small $\langle n_i \rangle$ and $\langle n_j \rangle$ one finds that the ratio of the Wehrl and the quantum mutual information scales as
\begin{align}
	\frac{J_{ij}^W}{I_{ij}} \approx \left[\ln \left( \frac{1}{\langle n_i \rangle \langle n_j \rangle} \right) \right]^{-1},
\end{align}
which goes to $0$ for $\langle n_i \rangle,{ }\langle n_j\rangle \rightarrow 0$. It should be noted that the Wehrl mutual information -- as the mutual information of the outputs of heterodyne measurements -- is only a lower bound for the classical correlation $J_M$. Nevertheless, since heterodyne measurements are often optimal~\cite{bradshaw2019, adesso2010, pirandola2014}, with a good level of certainty we can assume that classical intermode correlations become small in the low density regime.

Summarizing, we have observed a similar behavior for fermions and bosons in the low density limit, which is however quite paradoxical: Although their transport properties converge to those of classical particles, the entropy production in both cases becomes dominated by quantum intermode correlations. This again illustrates the lack of a direct relationship between the character of the reduced dynamics and the nature of microscopic correlations~\cite{pernice2012, smirne2021}.

	\section{Conclusions} \label{sec:concl}
	
	We conclude that the answer to the question whether the entropy production is mostly dominated by classical or quantum correlations is not universal but rather depends on the physical system. In particular, for bosonic systems one observes a quantum-to-classical transition in the microscopic nature of entropy production: it becomes dominated by the classical position-momentum correlations when the temperature increases. In contrast, no such transition is observed for fermionic systems, where correlations are always predominantly quantum. We relate this qualitative difference to the parity superselection rule applying to fermionic systems, which constrains the set of allowed projective measurements to projections on the Fock states, which limits the amount of classically accessible correlations. In contrast, no such rule applies to bosonic systems, enabling one to access a higher amount of correlations by performing Gaussian heterodyne measurements.
	
	We further note that the microscopic nature of the entropy production cannot be directly related to the character of the reduced dynamics. For example, the quantum contribution may be dominant even when the reduced dynamics and thermodynamics of the system can be effectively described by a Markovian master equation and the framework of stochastic thermodynamics~\cite{seifert2012}. Moreover, quite paradoxical behavior is observed in the low density limit when the Fermi and Bose-Einstein distributions converge to the Maxwell-Boltzmann distribution: While the transport properties of fermions and bosons become equivalent to those of classical particles, the entropy production is dominated by quantum correlations (also in the bosonic case). This lack of a direct relation between reduced and microscopic dynamics resembles similar observations for the pure dephasing case~\cite{pernice2012, smirne2021}, as well as our previous result showing that correlations generated within the environment are a dominant contribution to the entropy production even when they can be neglected in the reduced description of the system dynamics within the Born approximation (which assumes that the environment is in thermal equilibrium at all times)~\cite{ptaszynski2019}.
	
	Our paper opens new perspectives on research on the nature of correlations in different physical setups. First, the obvious direction is to consider coupled fermion-boson systems, which are ubiquitous in condensed matter physics~\cite{giustino2017} (e.g., in the context of superconductivity~\cite{bardeen1957}) and nanoscopic transport~\cite{koch2005, koch2006}; the particularly interesting question may be the nature of correlations between fermionic and bosonic degrees of freedom. Second, one can investigate other types of environment, such as spin baths~\cite{prokofev2000, breuer2004, cywinski2009, salamon2022}; in particular, one might expect a quantum-to-classical transition when increasing the spin dimension. The other objects of interest may be the systems of either topological~\cite{wilczek1982, wilczek1984, nayak2008} or statistical~\cite{myers2021} anyons -- quasiparticles continuously interpolating between bosons and fermions~\cite{carrillo2022}. Finally, as noted in Sec.~\ref{sec:fermnumsecor}, while we define the classical contribution to the entropy production as a correlation of outputs of local measurements on the system and all modes of the environment, possibly a higher amount of information can be accessed by enabling nonlocal measurements on sets of several modes. However, calculating such a quantity would be computationally demanding.
	
	\begin{acknowledgments}
K. P. has been supported by the National Science Centre, Poland, under Project No. 2017/27/N/ST3/01604, and by the Scholarships of Minister of Science and Higher Education. This research was also supported by the FQXi foundation Project No. FQXi-IAF19-05-52 ``Colloids and superconducting quantum circuits''.
	\end{acknowledgments}
	
	\appendix
	
	\section{Derivation of Eq.~\eqref{boundim}} \label{sec:hsnorm}
	To make the paper self-contained, in this section we repeat the derivation of Eq.~\eqref{boundim} presented by Bernigau \textit{et al}.~\cite{bernigau2013}. We first use the fact that the entropy production is bounded from below by the total correlation: $\sigma \geq I_M$. Furthermore, we write the formula for the von Neumann entropy [Eq.~\eqref{vnmcorf}] in the form $S=\text{Tr} s(\mathcal{C})$ where
	\begin{align}
		s(x)=-x \ln x -(1-x) \ln (1-x).
	\end{align}
	Thus, the total correlation can be calculated as $I_M=\text{Tr} s(\mathcal{C}^D)-\text{Tr} s(\mathcal{C})$. We then define the functions $l(x)=2x(1-x)$ and $g(x)=s(x)-l(x)$. Now, one can verify that all functions $s(x)$, $l(x)$ and $g(x)$ are concave in the whole range of eigenvalues $[0,1]$ of $\mathcal{C}$ and $\mathcal{C}^D$. In the next step, we use the Peierls inequality, which states that for any $N \times N$ Hermitian matrix $A$ and a function $f(A)$ that is concave within the range of eigenvalues of $A$ 
	\begin{align}
		\sum_{i=1}^N f(A_{ii}) \geq \text{Tr} f(A).
	\end{align}
	Thus, in particular,
	\begin{align}
		\text{Tr} g(\mathcal{C}^D) = \sum_{i=0}^K g(\mathcal{C}_{ii}) \geq \text{Tr} g(\mathcal{C}),
	\end{align}
	and therefore
	\begin{align}
		&I_M = \text{Tr} s(\mathcal{C}^D)- \text{Tr} s(\mathcal{C}) \geq \text{Tr} l(\mathcal{C}^D)-\text{Tr} l(\mathcal{C}) \\ \nonumber
		&= 2 \text{Tr} [\mathcal{C}^D (\mathds{1}_{K+1}-\mathcal{C}^D)]-2 \text{Tr} [\mathcal{C} (\mathds{1}_{K+1}-\mathcal{C})] = 2 \epsilon^2 \text{Tr} (\mathcal{E}^2),
	\end{align}
	where in the last step we use $\mathcal{C}=\mathcal{C}^D+\epsilon \mathcal{E}$ and $\text{Tr} \mathcal{E}=\text{Tr} (\mathcal{C}^D \mathcal{E})=0$; this proves Eq.~\eqref{boundim}.	
	
	We note that the above inequality is different from (and usually tighter than) the bound derived by Gullans and Huse~\cite{gullans2019}
	\begin{align}
		I_M \geq \frac{\epsilon^4 (K+1) \text{Tr} (\mathcal{E}^2)^2}{\langle \hat{N}_{SE} \rangle (K+1-\langle \hat{N}_{SE} \rangle)},
	\end{align}
	where $\langle \hat{N}_{SE} \rangle$ is the average total particle number. Furthermore, as shown in Ref.~\cite{gullans2019}, the bound $I_M \geq 2 \epsilon^2 \text{Tr} (\mathcal{E}^2)$ becomes tight for the infinite temperature state (with $\mathcal{C}^D=\mathds{1}_{K+1}/2$) in the limit of small $\epsilon$.
	
	\section{Derivation of Eqs.~\eqref{cormatcond0}--\eqref{cormatcond1}} \label{sec:cormatcond}
	To derive Eqs.~\eqref{cormatcond0}--\eqref{cormatcond1} we use the fact that a generic conditional state given output $\nu$ corresponding to the projective measurement $\Pi^\nu$ [obeying $\Pi^\nu=(\Pi^\nu)^\dagger$ and $\Pi^\nu \Pi^\nu=\Pi^\nu$] can be written as~\cite{nielsen2010}
	\begin{align}
		\rho^\nu=\Pi^\nu \rho \Pi^\nu/p_\nu,
	\end{align}
	where
	\begin{align}
		p_\nu=\text{Tr} (\Pi^\nu \rho)
	\end{align}
is the probability of the output $\nu$. Let us now consider the projective measurement on the occupied state of the mode $k+1$, which reads $\Pi_{k+1}^1=c^\dagger_{k+1} c_{k+1}$. The output probability is given by the average occupancy
\begin{align}
\text{Tr} (c_{k+1}^\dagger c_{k+1} \rho_{SE})=\mathcal{C}_{k+1,k+1}=\langle \hat{N}_{k+1} \rangle.
\end{align}
The elements of the conditional correlation matrix can then be calculated as
\begin{align} \nonumber
	\mathcal{C}^1_{ij} = &\mathcal{C}^{-1}_{k+1,k+1}\text{Tr} \left(c_i^\dagger c_j c_{k+1}^\dagger c_{k+1} \rho_{SE} c_{k+1}^\dagger c_{k+1} \right)= 
	\\ \nonumber & \mathcal{C}^{-1}_{k+1,k+1}\text{Tr} \left(c_i^\dagger c_j c_{k+1}^\dagger c_{k+1} \rho_{SE} \right)= 
	\\ \nonumber &\mathcal{C}^{-1}_{k+1,k+1} \left(\mathcal{C}_{ij} \mathcal{C}_{k+1,k+1}-\mathcal{C}_{i,k+1} \mathcal{C}_{k+1,j} \right)= 
	\\ &\mathcal{C}_{ij}-\mathcal{C}^{-1}_{k+1,k+1} \mathcal{C}_{i,k+1} \mathcal{C}_{k+1,j},
\end{align}
which gives Eq.~$\eqref{cormatcond1}$; here in the second step we use the cyclic property of the trace and fermionic anticommutation relations, while in the third step, we use Wick's theorem~\eqref{wick}. Eq.~$\eqref{cormatcond0}$ can be derived analogously using $\Pi_{k+1}^0=c_{k+1} c_{k+1}^\dagger$.

	\section{Calculation of Fock state probabilities} \label{sec:fockprob}
	Here we briefly describe the formalism used to calculate the Fock state probabilities $p(\mathbf{n})$ for $N$--mode bosonic Gaussian states developed by Dodonov \textit{et al.}~\cite{dodonov1994}. Within this approach, one first defines the matrix
	\begin{align}
		\mathcal{R} &=\mathcal{U}^* \left(\mathds{1}_{2N}-2 \Sigma \right) \left(\mathds{1}_{2N}+2 \Sigma \right)^{-1} \mathcal{U}^\dagger,
	\end{align}
	and the vector
	\begin{align}
		\mathbf{y} =2 \mathcal{U} \left (\mathds{1}_{2N} -2 \Sigma \right)^{-1} \mathbf{\bar{x}},
	\end{align}
	where
	\begin{align} \label{rotmat}
		\mathcal{U}=\frac{1}{\sqrt{2}} \begin{pmatrix} \mathds{1}_N & i \mathds{1}_N \\ \mathds{1}_N & -i \mathds{1}_N \end{pmatrix}.
	\end{align}
	Then one defines the generating function of Hermite polynomials
	\begin{align} \label{hermgen}
		\exp \left(-\frac{1}{2} \pmb{\gamma} \mathcal{R} \pmb{\gamma}^T + \pmb{\gamma} \mathcal{R} \mathbf{y}^T \right) =\sum_{\mathbf{m},\mathbf{n}=\mathbf{0}}^\infty \frac{\pmb{\beta}^{\mathbf{m}} \pmb{\alpha}^{\mathbf{n}}} {\mathbf{m}! \mathbf{n}!} H^{\{\mathcal{R}\}}_{\mathbf{m},\mathbf{n}}(\mathbf{y}),
	\end{align}
	where $\mathbf{n}!=\prod_{i=1}^N {n_i!}$, $\sum_{\mathbf{n}=\mathbf{0}}^\infty=\sum_{n_1=0}^\infty \ldots \sum_{n_N=0}^\infty$, $\pmb{\gamma}=(\beta_1,\ldots,\beta_N,\alpha_1,\ldots,\alpha_N)$, and $\pmb{\alpha}^\mathbf{n}=\prod_{i=1}^N \alpha_i^{n_i}$; in general, the parameters $\alpha_i$ and $\beta_i$ correspond to the complex amplitudes of coherent states, but for the sake of calculation of $p(\mathbf{n})$ they can be taken to be real. Finally, $H^{\{\mathcal{R}\}}_{\mathbf{m},\mathbf{n}}(\mathbf{y})$ are multidimensional Hermite polynomials which can be related to the state probabilities as
	\begin{align} \label{hermpol}
		p(\mathbf{n})=P_0 \frac{H^{\{\mathcal{R}\}}_{\mathbf{n},\mathbf{n}}(\mathbf{y})}{\mathbf{n}!},
	\end{align}
	where $P_0=p(0,\ldots,0)$ is the probability of the vacuum state:
	\begin{align}
		P_0=\left[ \det \left(\Sigma+\frac{\mathds{1}_{2N}}{2} \right) \right]^{-\frac{1}{2}} \exp \left[-\mathbf{\bar{x}} (2 \Sigma+\mathds{1}_{2N})^{-1} \mathbf{\bar{x}} ^T \right].
	\end{align}
	
	Using Eqs.~\eqref{hermgen}--\eqref{hermpol} the Fock state probabilities can be calculated as
	\begin{align}
		p(\mathbf{n})=\frac{P_0}{\mathbf{n}!} \left [ \frac{\partial^\mathbf{n}}{\partial \pmb{\alpha}^\mathbf{n}} \frac{\partial^\mathbf{n}}{\partial \pmb{\beta}^\mathbf{n}} 	\exp \left(-\frac{1}{2} \pmb{\gamma} \mathcal{R} \pmb{\gamma}^T + \pmb{\gamma} \mathcal{R} \mathbf{y}^T \right) \right]_{\pmb{\alpha},\pmb{\beta}=\mathbf{0}},
	\end{align}
	where $\pmb{\alpha}=(\alpha_1,\ldots,\alpha_N)$, $\mathbf{0}=(0,\ldots,0)$, and
	\begin{align}
		\frac{\partial^\mathbf{n}}{\partial \pmb{\alpha}^\mathbf{n}}=\frac{\partial^{n_1}}{\partial \alpha_1^{n_1}} \ldots \frac{\partial^{n_N}}{\partial \alpha_N^{n_N}}.
	\end{align}

	\section{Two-mode bosonic correlations for other symmetries of covariance matrix elements} \label{sec:bos2modesym}
\begin{figure}
	\centering
	\includegraphics[width=0.9\linewidth]{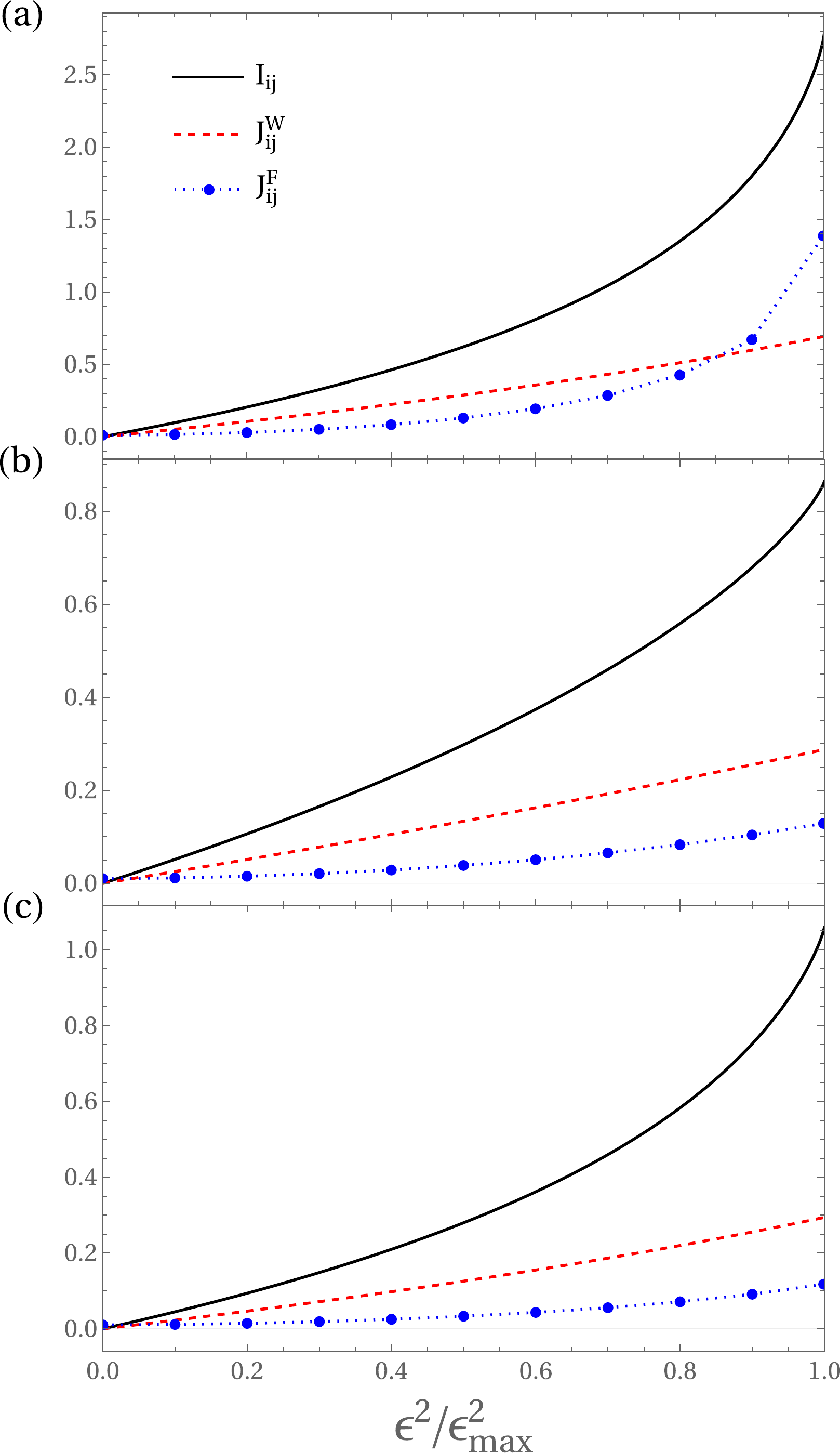}
	\caption{The total mutual information between two bosonic modes $I_{ij}$, the Wehrl mutual information $J^W_{ij}$, and the classical mutual information in the Fock basis $J_{ij}^F$ as a function of $\epsilon^2$ for $\langle n_i \rangle=\langle n_j \rangle=1$, and (a) $\epsilon_q=-\epsilon_p=\epsilon$, (b) $\epsilon_q=\epsilon_p=\epsilon$, and (c) $\epsilon_q=\epsilon$, $\epsilon_p=0$.}
	\label{fig:bos2s-lowoc}
\end{figure}

In Sec.~\ref{sec:bos2mode} we analyzed the intermode correlations for a two-mode bosonic system in the two-mode squeezed thermal state with $\epsilon_q=-\epsilon_p=\epsilon$. Here we investigate whether the results change for other choices of covariance matrix elements. Specifically, we focus on the case of $\epsilon_q=\epsilon_p$ and $\epsilon_q=\epsilon$, $\epsilon_p=0$. The maximum values of correlation terms read then as $\epsilon_\text{max}=\sqrt{n_1 n_2}$ and $\epsilon_\text{max}=2\sqrt{n_1 (1+n_1) n_2 (1+n_2)/[(1+2n_1)(1+2n_2)]}$, respectively. We found that for the latter case the calculation of Fock state probabilities for large occupancies becomes unfeasible (or, at lest, very time consuming) using a desktop computer. Therefore, we consider the case of a relatively low occupancy $\langle n_i \rangle=\langle n_j \rangle=1$ with a cutoff $n_j,n_j \leq 10$.

The results for different covariance matrix element symmetries are presented in Fig.~\ref{fig:bos2s-lowoc}. As one may observe, for all cases the Wehrl mutual information is dominant over the mutual information in the Fock basis in the regime of small $\epsilon$. The difference between different symmetries becomes only apparent in the high correlation regime of $\epsilon \approx \epsilon_\text{max}$. Then the two-mode squeezed thermal state becomes pure for $\epsilon=\epsilon_\text{max}$, and the mutual information in the Fock basis becomes dominant over the Wehrl mutual information. In contrast, for the other symmetries considered, the state remains mixed and $J^W_{ij}$ stays to be dominant.
	
	\section{Time evolution of the covariance matrix} \label{sec:covev}
	Here we present the method used to simulate the time evolution of the covariance matrix. We note that, as shown by Ullersma~\cite{ullersma1966}, the time evolution of the covariances can be expressed by means of (rather tedious) analytic equations; this approach has been applied, e.g., in Refs.~\cite{pucci2013, einsiedler2020, colla2021}. Here we use another technique, which is more clear and convenient for numerical implementation. Within this method, one considers an alternative form of the covariance matrix $\Sigma^C$ of the $N$--mode system with the matrix elements~\cite{adesso2014}
	\begin{align}
		\Sigma_{ij}^C= \frac{1}{2} \langle \{d_i,d_j \} \rangle - \langle d_i \rangle \langle d_j \rangle,
	\end{align}
	where $d_i=a_i$ and $d_{i+N}=a_i^\dagger$ ($i=1,\ldots,N$). The conversion between the covariance matrices reads
	\begin{align}
		\Sigma^C = \mathcal{U} \Sigma \mathcal{U}^\dagger,
	\end{align}
	with $\mathcal{U}$ defined in Eq.~\eqref{rotmat}. Any quadratic Hamiltonian can be further expressed in the form
	\begin{align}
		\hat{H}&=\frac{1}{2} \mathbf{d}^\dagger \mathcal{H} \mathbf{d}^T,
	\end{align}
	where $\mathbf{d}=(d_0,\ldots,d_{2N})$ and $\mathcal{H}$ is $2N \times 2N$ Hermitian matrix. In particular, for the Caldeira-Leggett Hamiltonian~\eqref{hamcald} one gets
\begin{align}
		\mathcal{H}_{SE}=\begin{pmatrix}
			\text{diag}(\pmb{\omega})-\mathbf{K}  & -\mathbf{K} \\ -\mathbf{K} & \text{diag}(\pmb{\omega})-\mathbf{K}
		\end{pmatrix},
	\end{align}
	where $\pmb{\omega}=(\omega_0,\ldots,\omega_K)$ and 
\begin{align}
    \mathbf{K}=\begin{pmatrix}
			0  & \pmb{\kappa} \\ \pmb{\kappa}^T & 0
		\end{pmatrix}
\end{align}
 with $\pmb{\kappa}=(\kappa_1,\ldots,\kappa_K)$. The evolution of the covariance matrix can be then expressed as~\cite{bruschi2021}
	\begin{align}
		\Sigma^C (t)=\mathcal{S} \Sigma^C(0) \mathcal{S}^\dagger,
	\end{align}
	where $\mathcal{S}=\exp(-i\mathcal{K} \mathcal{H}t)$ and
	\begin{align}
		\mathcal{K} = \begin{pmatrix}
			\mathds{1}_{N} & 0 \\ 0 & -\mathds{1}_N
		\end{pmatrix}.
	\end{align}

\end{document}